\documentclass[aps,prd,superscriptaddress,nofootinbib,floatfix]{revtex4}

\usepackage{slashed}
\usepackage{amsmath,amsfonts,amssymb}
\usepackage{booktabs}
\usepackage{comment}
\usepackage{siunitx}
\usepackage{verbatim}
\usepackage[english]{babel}
\usepackage{adjustbox}
\usepackage{relsize} 
\usepackage{makecell}

\usepackage{scalerel,stackengine}
\usepackage{titlesec}
\usepackage{parskip}
\usepackage{stix}
\usepackage{amssymb}
\usepackage{amsmath}
\usepackage{amsfonts}
\usepackage{hhline}
\usepackage{mathtools}
\usepackage[dvipsnames]{xcolor}
\usepackage{tcolorbox}

\usepackage{pifont}

\usepackage{multirow,tabularx}
\usepackage{siunitx}
\usepackage{multirow}
\usepackage{graphicx}
\usepackage{xstring}
\usepackage{etoolbox}
\usepackage{notoccite}
\usepackage{natbib}
\usepackage{mathrsfs}
\usepackage{lineno}
\usepackage{tensor}
\usepackage{accents}
\usepackage{tikz}

\usepackage{float}
\usepackage{tikz}

\usepackage{tikz-feynman}
\tikzfeynmanset{compat=1.1.0}

\usepackage{graphicx}

\newcommand{\be}{\begin{equation}}
\newcommand{\ee}{\end{equation}}
\newcommand{\bea}{\begin{equation}\begin{aligned}}
\newcommand{\eea}{\end{aligned}\end{equation}}

\def\nn{\nonumber}

\newcommand{\gsim}{\lower.7ex\hbox{$\;\stackrel{\textstyle>}{\sim}\;$}}
\newcommand{\lsim}{\lower.7ex\hbox{$\;\stackrel{\textstyle<}{\sim}\;$}}

\usepackage{hyperref}
\hypersetup{%
     colorlinks = true,%
     linkcolor = blue,%
     citecolor = blue,%
     filecolor = blue,%
     urlcolor = blue%
     }%
\usepackage[capitalize]{cleveref} 

\begin{document}

\title{Can MAG be a predictive EFT?\\
Radiative Stability and Ghost Resurgence in Massive Vector Models}

\author{Carlo Marzo}
\affiliation{Laboratory for High Energy and Computational Physics, NICPB, R\"{a}vala 10, Tallinn 10143, Estonia}
\date{\today}

\begin{abstract}
The rigorous conditions to obtain sensible predictions in non (proper) renormalizable Quantum Field Theories were derived a long time ago, most notably in the works of Steven Weinberg.
In this paper we explicitly illustrate the challenges met in carrying this program within the Affine Gravity framework, in particular when attempting to pinpoint viable particle propagation. We explore the one-loop structure of some ghost and tachyon-free vector theories to illustrate the role of structural constraints in their interactions, even in the absence of gauge symmetries.
Despite the presence of soft-breaking terms, we show how hopes of casting a vector model within the predictive frame of effective field methods hinges on adopting a gauge-like treatment of their interactions.
 \end{abstract}

\maketitle

\vspace{0.3cm}


\section{Introduction}

Effective Field Theory (EFT) offers a modern understanding of the remarkable phenomenological achievements of quantum field models \cite{Weinberg:1978kz,Donoghue:2012zc,Burgess:2007pt,Bijnens:2006zp,Burgess:2003jk,Colangelo:2001df,Manohar:1996cq,Kaplan:1995uv,Pich:1995bw,Ecker:1994gg,Donoghue:1994dn,Georgi:1993mps,Gasser:1984gg,Gasser:1983yg,Gasiorowicz:1969kn,Dobado:1989ax,Dobado:1989gr,Dobado:1990zh}. Focusing on the degrees of freedom of relevance for a given range of energy and momentum, it organizes their most generic dynamic through basic requirements such as causality, locality and Lorentz invariance. Because of its general approach, which allows for an infinite tower of non-renormalizable operators, it is expected to reproduce the low-energy behaviour of the unknown Physics lying on a more fundamental level.  

When this approach is supported by the careful identification of the infrared (IR) non-analytical effects, discerned from the local ultraviolet (UV) ones connected with the unknown Physics, unambiguous predictions can be made. Oversimplifying, the main differences between this set of \emph{Phenomenological Lagrangians} \cite{Weinberg:1978kz} and the narrower case of the renormalizable ones (asymptotically-free Yang-Mills being the archetype) are two. The first is the presence, in its defining way of organizing the infinite tower of non-renormalizable operators, of a cut-off scale signalling the limits of its applicability. The second is the need for a growing set of independent measurements to get increasingly precise predictions. 

Accounting for its virtues and limits, EFT has generated a vast range of low-energy theorems \cite{Weinberg:1978kz,Colangelo:2001df,Donoghue:1994dn,Dobado:1989ax,Dobado:1989gr,Dobado:1990zh,Donoghue:2015hwa} extending our knowledge beyond the reach afforded by the exclusive use of renormalizable quantum field theories (QFT). 
In this regard, the use of EFT methods for the quantum description of the gravitational interaction has been particularly insightful. The relevance of universal IR radiative profiling goes beyond the (admittedly optimistic) phenomenological implications, defining also the universality class for any realistic proposal for alternative theories of quantum gravity.   
Key to this success is the use of a gauge symmetry, diffeomorphism invariance, which forbids relevant and marginal operators \footnote{Here, marginal, relevant and irrelevant refer to the operator mass-dimension in natural units.} and tightly constraint the form of the expected UV divergences. The imposition of these structural constraints over the counterterm Lagrangian is crucial in ensuring a predictive application of the EFT program \cite{Gomis:1995jp}. Moreover, the same symmetry controls also the possible detuning of the kinetic term. This is a concern for the potential generation of Ostrogradsky instabilities \cite{Ostrogradsky:1850fid,Woodard:2015zca}, as discussed in \cite{Simon:1990jn,Anselmi:2002ge}, connected to high-derivatives in bilinear operators. When the protecting symmetries are lacking altogether, Ostrogradsky instabilities can also be accompanied by further transgressions when ghostly and/or tachyon components, tuned away in the starting Lagrangian, are generated by radiative corrections. 
In light of this successful EFT treatment, the pursuit to extend the program to the framework of Metric Affine Gravity has enjoyed intriguing progress, in particular through the classifications and analyses of \cite{PhysRevD.56.7769,Baldazzi:2021kaf,Pradisi:2022nmh}. 
Metric Affine Gravity (MAG) \cite{Hehl:1994ue,BeltranJimenez:2019acz,Iosifidis:2023but,Shapiro:2001rz,Mondal:2023cxx,Rigouzzo:2023sbb,Rigouzzo:2022yan,Karananas:2021zkl,Karananas:2018nrj,
Neville:1978bk,Sezgin:1979zf,Sezgin:1981xs,Karananas:2014pxa,Lin:2018awc,Lin:2020phk,Neville:1979rb,Mikura:2024mji,Mikura:2023ruz,BeltranJimenez:2019hrm,
Annala:2022gtl,Barker:2024ydb,Barker:2024dhb}, which widens the gravitational field content with an independent rank-3 tensor, naturally aggravates the potential for disruptive propagation. 
At tree level, the search for ghost and tachyon-free MAG models still draws considerable attention, as naturally expected, given the large number of particle components carried by unconstrained three-index tensor. After such a first stage of selection for viable linear models, it is essential to enquire about the structure of the radiative corrections introduced.   
This has been recently tackled \cite{Melichev:2023lwj,Baldazzi:2021kaf}, continuing the past efforts of \cite{YuBaurov:2018pyj,Buchbinder:1985jc,Kalmykov:1994fm}. As expected, the radiative generation of new, high-order operators, while manifesting non-renormalizability in the Dyson sense, leaves the door open for an EFT treatment of MAG, thus enabling a predictive phenomenological use. 

In this paper, the first of two dedicated to the subject of Phenomenological Lagrangians in MAG, we aim to highlight the main obstructions, and point to possible solutions in seeking a successful EFT framing of MAG. In turn, we leave the presentation of models successfully defining predictive propagation to the forthcoming second part. As a case study, we consider here the healthy linear propagation of a massive vector so to build a parallel with the traditional rank-1 effective descriptions \cite{Coates:2022qia,Coates:2023swo,Clough:2022ygm,Mou:2022hqb,East:2022ppo,Heisenberg:2014rta,Heisenberg:2016eld,BeltranJimenez:2016rff,Heisenberg:2020jtr,Barker:2023fem}. 

\section{Effective Theories of Massive Vectors}
There is no clearer display of the lurking perils proper to high-rank model-building than the study of the candidate interactions of a massive vector field. To those aware that the dynamic of a massive vector can be successfully framed in U(1)-invariant descriptions, the proposed interactions will cause a natural scepticism. This is, indeed, the correct reaction in recognizing that, while there are no gauge symmetries in the free theory, they are needed to ensure that interactions do not affect the dominant non-perturbative propagation. This attitude, as we hope to illustrate via explicit examples, should be transposed to any high-rank (effective) interacting theory where less experience with working models and corresponding predictions has been developed. 
The concerns, of a very practical nature, will remain the same for every high-rank field model: 
\begin{itemize}
   \item  {\textsc{Modern (non-)renormalization:}} \emph{Structural constraints are needed to ensure that counterterms exist for all the divergences to be, at each order, absorbed by renormalizing amplitudes at a given scale. Theories with unconstrained coefficients set to zero, at all scales, are therefore inconsistent \cite{Gomis:1995jp}.}
  \item {\textsc{Large momentum behaviour of the propagator:}} \emph{Even when there are no explicit gauge symmetries, due to soft mass terms, the generation of momentum dependence along the ghost-like components needs to be either completely nullified or shifted outside the range of validity of a viable effective description.}
   \end{itemize}
While the first statement almost serves as a foundational pillar for reasonable EFT studies, the second, while quite valuable, is not a sufficient condition to get a predictive theory. 
The presence of symmetry-breaking soft terms is indeed enough to not introduce \emph{local} (UV-induced) instabilities, but their finite imprints can still spoil some EFT expectations and ask for a case-by-case study. For Proca/Abelian theories, the soft terms (as the explicit mass) can indeed be made completely harmless and we expect this necessary condition to also become sufficient.

\subsection{Proca, bottom-up: Self-Interaction and Ghostly Scalars} \label{ProcaWeird}
The Proca quadratic Lagrangian emerges from the simplest inclusion of unitarity and tachyon-freedom for a rank-1 field $A_{\mu}(x)$. We use the customary succinct notation $S_n^p$ to represent the n-th massive particle representation of spin S and parity p, carried by a given Lorentz tensor \cite{Percacci:2020ddy}. 
We will not repeat here the very simple spectral analysis but remind the reader that the massive propagations of the three-component $1^-$ and the scalar $0^+$ sectors carried by the Lorentz vector cannot be simultaneously healthy. 
Consequently, to propagate a massive vector, we are left with the choice 
\\
\begin{align} \label{ProcaAc}
&\mathcal{S}_2  = - \frac{1}{2} \int \frac{d^4 p}{(2\pi)^4} \,A^{\mu}(p)\,\left( g_{\mu \nu} \left(p^2 - m_V^2\right) - p_{\mu}p_{\nu}\right) A^{\nu}(-p) = - \frac{1}{2} \int \frac{d^4 p}{(2\pi)^4} \, A^{\mu}(p)\,\left( P^{1^-}_{\mu \nu} \left(p^2 - m_V^2\right) -  m_V^2 P^{0^+}_{\mu \nu}\right) A^{\nu}(-p) \nn \\ 
& \hspace{0.5cm} = \frac{1}{2}  \int d^4 x  A^{\mu}(x) \left(g_{\mu \nu} (\Box + m_V^2) - \partial_{\mu} \partial_{\nu}\right) A^{\nu}(x) \,, \hspace{1.cm} \left(\text{with}\hspace{0.4cm} P^{1^-}_{\mu \nu} = g_{\mu \nu} - \frac{p_{\mu} p_{\nu}}{p^2} \, ,\,\, P^{0^+}_{\mu \nu} = \frac{p_{\mu} p_{\nu}}{p^2}\, \right)
\end{align}
that is, Proca theory. The lengthy form adopted in eq.~\ref{ProcaAc} shows explicitly that all the particle sectors (accessed to via the corresponding projectors $P^{0^+}$ and  $P^{1^-}$) appear in the quadratic Lagrangian, thus no obstructions to the computation of the propagator are generated and the theory is devoid of gauge symmetries. For this reason, self-interacting Proca theories have often welcomed terms polynomial in $A^2 = A_{\mu} A^{\mu}$ \cite{Coates:2022qia,Coates:2023swo,Clough:2022ygm,Mou:2022hqb,East:2022ppo,Heisenberg:2014rta,Heisenberg:2016eld,BeltranJimenez:2016rff,Heisenberg:2020jtr,Hell:2021oea,Hell:2021wzm}. While such theories are perturbatively non-renormalizable due to the (Euclidean) large-momentum behaviour of the Proca propagator 
\begin{align}
    D_{\mu \nu} = \frac{- i }{q^2 - m_V^2} \left(g_{\mu \nu} - \frac{q_{\mu}q_{\nu}}{m_V^2} \right) =  - i \left( \frac{P^{1^-}_{\mu \nu}}{q^2 - m_V^2} -  \frac{P^{0^+}_{\mu \nu}}{m_V^2}\right) \, , 
\end{align}
we might enquire if they can effectively describe massive vector interactions in some predictive EFT framework. 
We consider the following non-linear addendum to eq.~\ref{ProcaAc}
\begin{align} \label{ProcaIn}
    \mathcal{S}_{i} = - \int d^4 x \,\bigg( \frac{g_3}{4}  A_{\nu}(x) A^{\nu}(x) \partial_{\mu}A^{\mu}(x) + \frac{g_4}{4} \left(A_{\nu}(x) A^{\nu}(x)\right)^2 \bigg)\, ,
\end{align}
which forms, for a unique abelian field, the most generic self-interacting setup with operators of dimension $d_O \leq 4$.
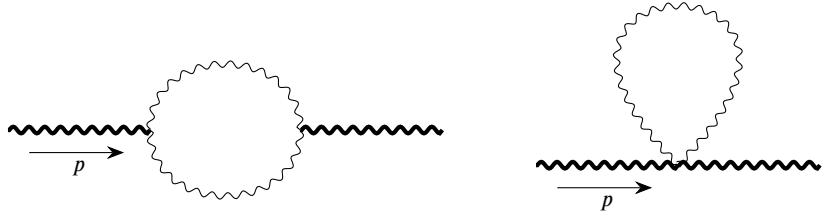
\begin{figure} 
\label{Fig:AA}
\centering
\begin{tikzpicture}
  \begin{feynman}
    \vertex[blob, red, minimum size=2cm] (AA) at (3,0);
    \vertex (a) at (0, 0) {};
    \vertex (b) at (2, 0);
    \vertex (c) at (4, 0);
    \vertex (d) at (6, 0) {};
   \diagram* {
      (a) -- [photon, black, ultra thick, momentum'={\(\color{black} p\)} ] (b) -- [half left, photon, black] (c),
      (c) -- [half left, photon, black] (b),
      (c) -- [photon, black, ultra thick] (d),
    };
  \end{feynman}
\end{tikzpicture}
\begin{tikzpicture}
  \begin{feynman}
    \vertex (a) at (0, 0) {};
    \vertex (b) at (2, 0);
    \vertex (t) at (2, -2);
    \vertex (d) at (4, 0) {};
   \diagram* {
   (a) -- [photon, black, ultra thick, momentum'={\(\color{black} p\)} ] (b) -- [out=135, in=45, loop, photon, black, min distance=4cm] b -- [photon, black, ultra thick] (d),
   };
  \end{feynman}
\end{tikzpicture}
\caption{Self-energy corrections induced by vector self-interaction. }
\end{figure}

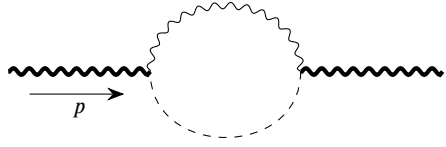
\begin{figure} 
\label{Fig:AASca}
\centering
\begin{tikzpicture}
  \begin{feynman}
    \vertex[blob, red, minimum size=2cm] (AA) at (3,0);
    \vertex (a) at (0, 0) {};
    \vertex (b) at (2, 0);
    \vertex (c) at (4, 0);
    \vertex (d) at (6, 0) {};
   \diagram* {
      (a) -- [photon, black, ultra thick, momentum'={\(\color{black} p\)} ] (b) -- [half left, photon, black] (c),
      (c) -- [half left, scalar, black] (b),
      (c) -- [photon, black, ultra thick] (d),
    };
  \end{feynman}
\end{tikzpicture}
%
\caption{Self-energy corrections induced by effective real scalar-vector interaction. }
\end{figure}
We can draw a clear picture of the difficulties introduced by eq.~\ref{ProcaIn} via the straightforward appraisal of the one-loop radiative contribution to the two-point function fig.~\ref{Fig:AA}. Simple manipulations reveal the following UV-singular structure in momentum-space (which we regularized computing in generic dimensions $D = 4 - 2\epsilon$)
\begin{align} \label{amplitude}
&\langle A_{\mu}(p) A_{\nu}(-p) \rangle \propto \frac{1}{\epsilon} \, \frac{g_3^2}{(4 \pi)^{2}}  \bigg[ \left(-\frac{3}{16} m_V^2 - \frac{3}{16} p^2\right) P^{1^-}_{\mu \nu}  +  \left(-\frac{3}{16} m_V^2 + \frac{9}{16} p^2 - \frac{3}{16}\frac{p^4}{m_V^2} + \frac{1}{32}\frac{p^6}{m_V^4} \right) P^{0^+}_{\mu \nu}\bigg]  + \nn \\
& \hspace{1cm} +  \frac{1}{\epsilon} \, \frac{g_4}{(4 \pi)^{2}}  \bigg[ \left(-\frac{9}{2} m_V^2\right) P^{1^-}_{\mu \nu}  +  \left(-\frac{9}{2} m_V^2 \right) P^{0^+}_{\mu \nu}\bigg]  + \text{UV-finite} \, . 
\end{align}
When expressed in terms of local, field-valued operators, the divergent components of the one-loop result call for the following setup\footnote{We adopted a one-to-one correspondence among operators and couplings to give a direct image of the renormalization procedure. This choice, while perfectly sound, hides that one combination, connected to field-rescaling, is an inessential parameter.} 
\begin{align} \label{RenAction}
    & \frac{1}{2}  \int d^4 x  A^{\mu}(x)\bigg[ \nn \\
    & \hspace{0.3cm} \left( Z^0_T + Z^2_T\frac{\Box}{m_V^2} + Z^4_T\frac{\Box^2}{m_V^4} + \cdots \right) \left(g_{\mu \nu} \Box - \partial_{\mu} \partial_{\nu}\right) + Z_m \,m_V^2 \,g_{\mu \nu} + \left( Z^0_L + Z^2_L\frac{\Box}{m_V^2} + Z^4_L\frac{\Box^2}{m_V^4} + \cdots  \right) \,\partial_{\mu} \partial_{\nu}  \bigg] A^{\nu}(x) \nn \\ 
\end{align}
where each of the $Z_i^j$ parameters can be split into a pole and a finite part
\begin{align} \label{RenCoupling}
Z_i^j = \frac{1}{(4\pi)^2 \epsilon} z_i^j + \tilde{Z}_i^j(\mu)\, ,
\end{align}
with the finite part connected to fixed values of the renormalized two-point amplitude. Notice how an explicit dependence from the unphysical renormalization scale, here symbolized by $\mu$, is the finite counterpart of the unphysical dependence from the regulator $\epsilon$.  
The direct computation gives, in a scale-less renormalization, the following values for the singular parts\\
\begin{align}
    & z_m = \frac{3}{16} \left(g_3^2  + 24 g_4^2\right) \,, \hspace{1cm} z_T^0 = - \frac{3}{16}  g_3^2\,, \hspace{1cm} z_T^2 =z_T^4 = 0\,, \hspace{1cm} z_L^0 =  \frac{9}{16} g_3^2\, , \hspace{1cm} z_L^2 = \frac{3}{16}  g_3^2\,, \hspace{1cm} z_L^4 = \frac{1}{32} g_3^2 \,. \nn
\end{align}
While the first two terms are a renormalization of the Lagrangian parameters in eq.~\ref{ProcaAc}, the remaining set presents a more upsetting scenario.
The uncurbed longitudinal components of the propagator, triggered by the cubic interaction in eq.~\ref{ProcaIn}, generate momentum dependence along $A^{\mu} \partial_{\mu}\partial_{\nu} A^{\nu}$. The presence of the UV-singular component will require a scale-dependent finite counterpart $\tilde{Z}_L^0(\mu)$ which, \emph{for any value}, will always provide a ghost-like state if the $1^-$ sector is healthy. Attempts to ignore the ghostly polarizations in the initial/final state of the scattering process will, unless properly prescribed \cite{Anselmi:2018kgz,Anselmi:2018tmf,Anselmi:2017yux,Piva:2023eaj,Piva:2023noi}, break the validity of the optical theorem, and therefore proceed to harass unitarity.  
Even ignoring the inconsistencies connected to the two-derivative longitudinal operator, the higher-order operators present a further obstacle to any practical use of such a theory. Differently from the predictive framework of EFT, the higher-order operators needed to renormalize the amplitude are not dampened by any large cut-off mass but scale accordingly with the only dimensionful parameter available, the mass of the vector. 

\begin{tcolorbox}[colback=lightgray, colframe=black, width=\textwidth, sharp corners]
   \emph{Being this also the scale of the relevant propagating particles of the theory, we cannot adopt in any case the usual double perturbative expansion, in small coupling and derivatives, as typical of functioning EFTs.}
\end{tcolorbox}

We reiterate this remark which is sometimes a source of confusion: \emph{$m_V$ is not the cut-off of the EFT,} being it part of the healthy states of the spectrum. Otherwise, the entire Lagrangian under investigation would be relegated to irrelevance, which is not the assumption made and contradicts the original search for ghost and tachyon-free propagation.  

Because of this, it is sometimes tempting to force the introduction of an extra large scale for the dimensionless set of couplings in eq.~\ref{ProcaIn}, as in 
\begin{align}
    g_3 \rightarrow \tilde{g}_3 \frac{m_V^2}{m_{\Lambda}^2} \, ,  \hspace{1cm}  g_4 \rightarrow \tilde{g}_4 \frac{m_V^2}{m_{\Lambda}^2} \,, 
\end{align}
so to recover an apparent EFT-like scaling of the UV-part of eq.~\ref{RenAction}. It is unclear to us if this procedure can push away the dangerous terms at all orders. Moreover, at the price of a one-loop control over the renormalization procedure, the rescaling forces the use of remarkably small dimensionless couplings which also dampen the infrared (IR) universal prediction of the computation. While this does not represent an inconsistency per se, it questions the reasons for the introduction of the dimensionless couplings in the first place, being such suppression, as we will see in Sec.~\ref{Sec:SIP}, the default of the standard approach to Proca EFTs based on U(1)-symmetric interactions.
As a final remark, we notice that much of the trouble is generated by the cubic interaction in eq.~\ref{ProcaIn}, while the quartic appears to not detune the quadratic Lagrangian. This is, in a way, a coincidence connected to the momentum-independence of the tadpole diagram in \ref{Fig:AA} and does not generalize to direct inspection of two-loop contributions Fig.~\ref{Fig:AA2L} which are well described by the operators of eq.~\ref{RenAction}.
\subsection{Effective Theory for a Interacting Proca Field} \label{Sec:SIP}
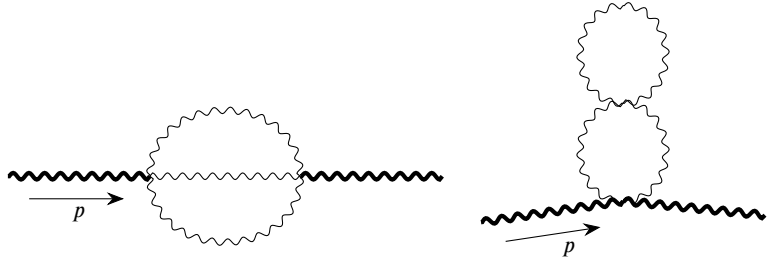
\begin{figure} 
\label{Fig:AA2L}
\centering
\begin{tikzpicture}
  \begin{feynman}
    \vertex[blob, red, minimum size=2cm] (AA) at (3,0);
    \vertex (a) at (0, 0) {};
    \vertex (b) at (2, 0);
    \vertex (c) at (4, 0);
    \vertex (d) at (6, 0) {};
   \diagram* {
      (a) -- [photon, black, ultra thick, momentum'={\(\color{black} p\)} ] (b) -- [half left, photon, black] (c),
      (c) -- [half left, photon, black] (b),
      (c) -- [photon, black, ultra thick] (d),
      (c) -- [photon, black] (b),
    };
  \end{feynman}
\end{tikzpicture}
\begin{tikzpicture}
  \begin{feynman}
    \vertex (a) at (0, -0.3) {};
    \vertex (b) at (2, 0);
    \vertex (t) at (2, 1.3);
    \vertex (tt) at (2, 2.6);
    \vertex (d) at (4, -0.2) {};
   \diagram* {
   (a) -- [photon, black, ultra thick, momentum'={\(\color{black} p\)} ] (b) -- [half left, photon, black] (t), 
   (t) -- [half left, photon, black] (b),
   (t) -- [half left, photon, black] (tt), 
   (tt) -- [half left, photon, black] (t),
   (b) -- [photon, black, ultra thick] (d),
   };
   \end{feynman}
\end{tikzpicture}
\caption{Self-energy two-loop corrections induced by vector self-interaction. }
\end{figure}
If we limit ourselves to the study of self-interacting abelian vectors we can consider, instead of eq.~\ref{ProcaIn}, the more modest selection of non-renormalizable interactions
\begin{align} \label{ProcaEff}
\mathcal{S}_{i} = - \int d^4 x \,\bigg( \frac{c_3}{4} \frac{(F^2)^2}{m_{\Lambda}^4}  + \frac{c_4}{4} \frac{Tr[F^4]}{m_{\Lambda}^4}  + \sum_{i}^{\infty} \lambda_i \, \omega_i \left[F,\partial/m_{\Lambda}\right]   \bigg)
\end{align}
where the infinite sum stands for polynomials, of increasing dimensionality, built out of space-time derivatives and the U(1)-invariant form $F_{\mu \nu} = \partial_{\mu} A_{\nu} - \partial_{\nu} A_{\mu}$. The large mass $m_{\Lambda}$ is now introduced to justify the neglecting of higher-order operators and enters naturally in the definition of the dimensionful couplings. The second diagram of \ref{Fig:AA} contributes only to the renormalization of the kinetic term, that is, the parameters $Z^0_T$ and $Z_m$ of eq.~\ref{RenAction} 
%
\begin{align} 
 & z_m = 0 \,, \hspace{1cm} z_T^0 =  - \frac{m_V^4}{m_{\Lambda}^4} \left(7 c_3 + 16 c_4\right)\,. 
\end{align}
Notice the appearance of the dimensionless ratio ${m_V}/{m_{\Lambda}}$ which accompanies all the perturbative contributions computed from eq.~\ref{ProcaEff}.
To infer the impact of the interactions on the vector propagation we need, therefore, to consider the higher-order terms of fig.~\ref{Fig:AA2L}.
The presence of the mass term does not spoil the U(1)-invariant structure of the counterterms and the dangerous longitudinal terms are never generated, giving $Z_L^i = 0$ at all orders. What the presence of the mass terms cannot guarantee is the absence of higher-order poles along the transversal direction.  A generic argument \cite{Anselmi:2002ge}\footnote{We are particularly thankful to Marco Piva for the guidance provided on this particular point.} shows that in a massless U(1)-invariant theory the divergences of the form 
\begin{align} \label{anselmi}
    F_{\mu \nu} \partial_{\rho_1} \cdots \partial_{\rho_{2 n}} F_{\alpha \beta} , \,  \hspace{1.cm}   F_{\mu \nu} \Box^n F^{\mu \nu}\, ,
\end{align}
can be reformulated in terms of vertex corrections through the combined use of Bianchi identities and the equations of motion
\begin{align} \label{eoms}
  \partial_{\mu} F_{\nu \sigma} + \partial_{\sigma} F_{\mu \nu} + \partial_{\nu} F_{\sigma \mu} = 0\, , \hspace{1.3 cm}  \partial^{\mu} F_{\mu \nu} = \mathcal{O}(F^2) \, . 
\end{align}
In the presence of a mass term, the equations of motion become 
\begin{align}
    \partial^{\mu} F_{\mu \nu} = m_V^2 A_{\nu} + \mathcal{O}(F^2) \, , 
\end{align}
and the complete reformulation of eq.~\ref{anselmi} in terms of higher-order vertices is no longer granted.  
Nevertheless, in a U(1)-invariant EFT as eq.~\ref{ProcaEff}, the UV renormalization of higher-order poles can only be triggered by non-renormalizable operators with coefficients proportional to negative powers of the large scale $m_{\Lambda}$. Consequently, the bare quadratic Action is now more aptly described in a form that embodies the invariance and the effective character of the higher-order corrections:
\begin{align} \label{RenActionEff}
   &  \frac{1}{2}  \int d^4 x  A^{\mu}(x)\bigg[\left( Z^0_T + Z^2_T\frac{\Box}{m_{\Lambda}^2} + Z^4_T\frac{\Box^2}{m_{\Lambda}^4} + \cdots + Z^i_T\frac{\Box^{i}}{m_{\Lambda}^{2 i}} \right) \left(g_{\mu \nu} \Box - \partial_{\mu} \partial_{\nu}\right) + Z_m \,m_{V}^2 \,g_{\mu \nu}  \bigg] A^{\nu}(x) = \nn \\ 
   &   \int d^4 x \left[- \frac{1}{4} \, F_{\mu \nu} \left( Z^0_T + Z^2_T\frac{\Box}{m_{\Lambda}^2} + Z^4_T\frac{\Box^2}{m_{\Lambda}^4} + \cdots + Z^i_T\frac{\Box^{i}}{m_{\Lambda}^{2 i}} \right) F^{\mu \nu}  + \frac{1}{2} Z_m m_V^2 A^{\mu}(x) A_{\mu}(x)\right]\, ,
\end{align}
so that the emerging higher-order poles are naturally connected with states of mass $\sim m_{\Lambda}$. These values sit outside the range of validity of our EFT ($\text{energy}/m_{\Lambda} \ll 1$), where effects from the fundamental, renormalizable theory become relevant. 
\\
Indeed, as can be inferred from the absence of one-loop higher-order poles, and a direct assessment of the two-loop diagrams Fig.~\ref{Fig:AA2L}, the first contribution to the transversal dipole $Z_T^2 F_{\mu \nu} (\Box/{m_{\Lambda}^2}) F^{\mu \nu}$ experiences a stronger suppression 
\begin{align} \label{TwoLoopFF}
    Z_T^2  \propto \frac{1}{\epsilon^2} \frac{m_V^6}{m_{\Lambda}^6}\, ,
\end{align}
in sharp contrast with the behaviour induced by eq.\ref{ProcaIn}. We emphasize that this particularly convergent/weak contribution is a prediction of the most generic EFT of a self-interacting abelian vector which includes the tower of interactions in eq.~\ref{ProcaEff}. This can be appreciated already by including the effective invariant interaction with a massless \emph{real} scalar $\phi(x)$   
\begin{align}
    \mathcal{S}_{\phi}  = - \int d^4 x \left( \frac{c_s}{m_{\Lambda}} F^2 \phi \right)\, ,
\end{align}
which would give, from the diagram in fig.~\ref{Fig:AASca}, the values
\begin{align} 
 & z_m = 0 \,, \hspace{1cm} z_T^0 =  - 8\, \frac{m_V^2}{m_{\Lambda}^2}  c_s^2 \,, \hspace{1cm} z_T^2 =  \frac{8}{3}  c_s^2\, . 
\end{align}

\begin{figure}[h] 
    \centering
    \includegraphics[width=0.6\textwidth]{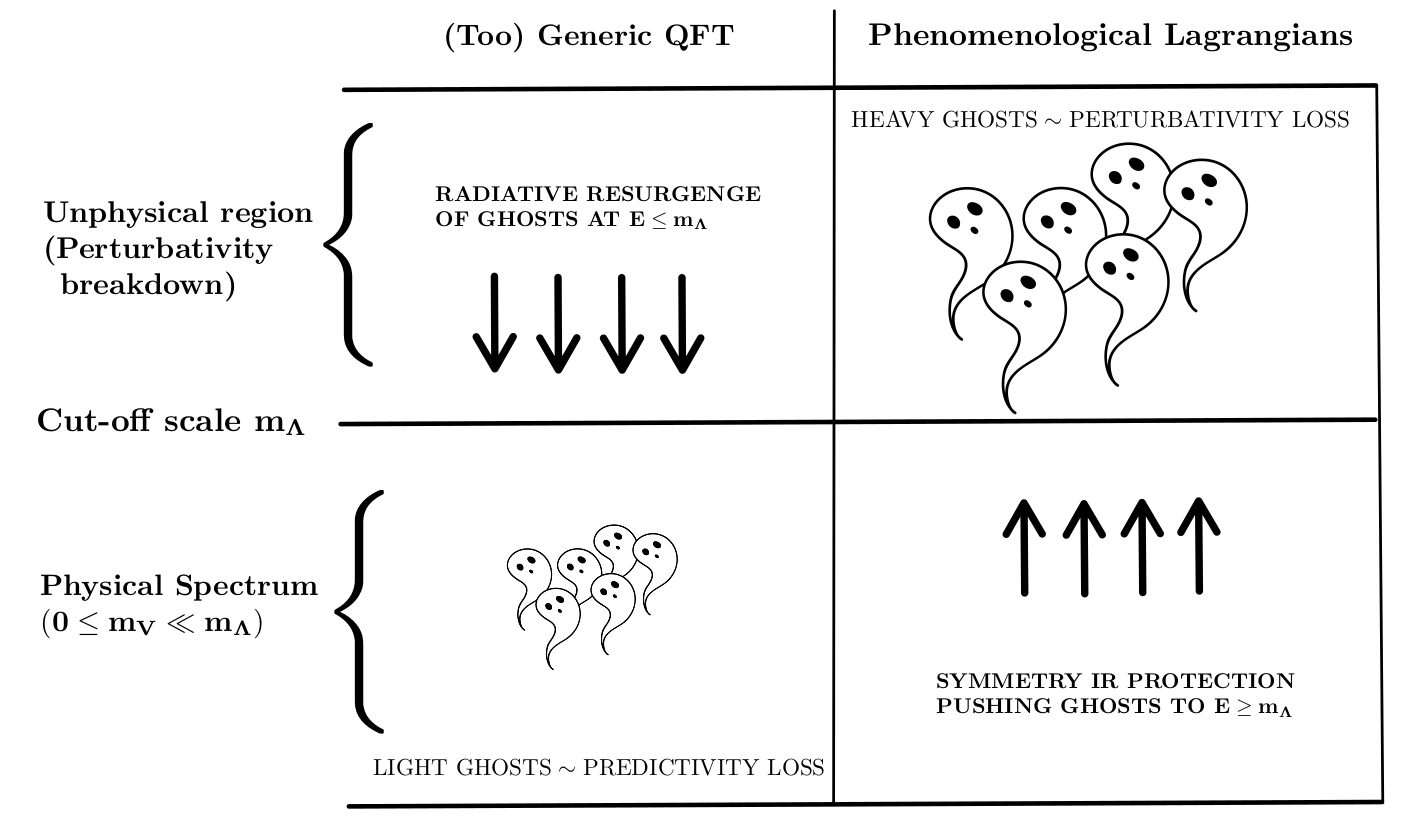} 
    \caption{Graphical representation of the effect of uncurbed model building with fields of rank $\geq$ 1. }
    \label{fig:ghosts}
\end{figure}

\section{Effective descriptions in MAG}
%
It is the focal point of our effort to illustrate that the very same issues afflicting careless modelling of vector interactions can appear, amplified, in MAG. We continue by examining the case of massive spin-1 propagation, this time carried by the rank-3 tensor $A_{\mu \nu \rho}(x)$ entering, as an independent field, the definition of the non-Riemann curvature tensor
\begin{equation} \label{curvature}
	F_{\,\mu\nu \,\, \sigma}^{\,\,\,\,\,\, \rho} \equiv 2\left(\partial_{[\mu} A_{\nu] \,\,\, \sigma}^{\,\,\,\,\rho} +  A_{[\mu| \,\,\, \alpha}^{\,\,\,\,\,\,\rho} A_{|\nu] \,\,\, \sigma}^{\,\,\,\,\,\,\alpha}\right).
\end{equation}
The challenges to oppose in building a predictive interacting theory for the affine connection $A_{\mu \nu \rho}(x)$ are twofold. One type is generically connected with the high-rank nature of the corresponding Lorentz tensor which will channel, if not precisely controlled, the propagation of both healthy as well as ghostly states. Even for our case study, the subset of two-index symmetric rank-3 tensors ($A_{\mu \nu \rho}  =  A_{\rho \nu \mu}$), the twelve particle sectors 
\begin{align} \label{ParticleDecT}
&A{_{\mu \nu \rho}} \supset \Big\{ 3_1^- , 2_1^+ , 2_2^+ , 2_1^- , 1^+_1 , 1^-_1,  1^-_2 ,	1^-_4,  1^-_5 , 0^+_1, 0^+_2, 0^+_4 \Big\},
\end{align}
are to be tamed within the boundaries of causal and unitary propagation. The other troubling feature is instead specific to MAG which uses, due to its geometrical bearing, the curvature tensor eq.~\ref{curvature} as the building block of invariant Lagrangians. 
This structure intertwines linear and quadratic terms, therefore connecting propagation and vertices in a Lagrangian built out of its possible invariant contractions.
This is often seen as a positive hint towards bridging MAG with the successful framework of Yang-Mills theory, which is also built from invariants (the plural to allow effective interactions) of its own curvature tensor 
\begin{align} \label{YM}
   F_{\mu \nu}^{i} \equiv 2 \partial_{[\mu} B_{\nu]}^{\,i} - g f^{i j k } B_{\mu}^{\,j} B_{\nu}^{\,j}  \, . 
\end{align}
This link is purely cosmetic. The shape of eq.~\ref{YM} is dictated by the consistent deformation of the gauge symmetry $\delta B^i_{\mu}(x) = \partial_{\mu} \epsilon^i(x)$ which removes, as explicitly checked in the previous section, the ghost-like longitudinal state from the spectrum. Conversely, the shift symmetry $\delta A_{\mu \,\,\, \sigma}^{\,\,\,\,\rho}(x) = \partial_{\mu} \partial_{\sigma} \xi^{\rho}(x) $ that shapes the curvature  eq.~\ref{curvature} \emph{is not} connected to any tuning within the components carried by $A_{\mu \,\,\, \sigma}^{\,\,\,\,\rho}$ (a subgroup of which is displayed in eq.~\ref{ParticleDecT}). It is, instead, directly connected with the protecting symmetry of the rank-2 symmetric tensor $h_{\mu \nu}$, which propagates the graviton as a massless spin-2 particle. Such a connection is forced by the use of the rank-3 tensor $A_{\mu \,\,\, \sigma}^{\,\,\,\,\rho} = A_{\sigma \,\,\, \mu}^{\,\,\,\,\rho}$ as a non-dynamical Lagrange multiplier in the first-order Palatini formulation of the Einstein-Hilbert theory
\begin{align} \label{Palatini}
 \mathcal{S}_{EP} = a_{EP} \int d^4 x \left[ h^{\mu \nu} \left( \partial_{\alpha} A_{\mu \,\,\, \nu}^{\,\,\,\,\alpha} - \partial_{\nu} A_{\mu \,\,\, \alpha}^{\,\,\,\,\alpha}\right) + A_{\,\,\mu}^{\alpha\,\,\mu}A_{\alpha \,\,\, \beta}^{\,\,\,\,\beta} - A_{\beta \,\,\, \mu}^{\,\,\,\,\alpha} A_{\alpha}^{\,\,\beta \mu} + \mathcal{O}(H A^2) \right]  \, .   
\end{align}
In other words, the shift symmetry of the rank-3 tensor is the protecting gauge symmetry of the \emph{graviton} field disguised through the use of first-order formalism. It plays no role in any sought cancellation when eq.~\ref{Palatini} becomes the starting point for promoting the affine connection $A_{\sigma \,\,\, \mu}^{\,\,\,\,\rho}$ to a dynamical field. 
It is a direct consequence of this mismatch another possible source of caution. Because of the connection between linear and quadratic terms in eq.~\ref{curvature}, in using the usual list of geometrically-justified quadratic invariants we (often) automatically generate a corresponding non-linear part. In such cases, for the cited reasons, many studies of ghost-free theories that focused only on the quadratic part of a candidate Lagrangian, automatically include a selection of vertices that, in the same ways illustrated for the Proca model of eq.~\ref{ProcaIn}, will detune the propagation radiatively.  \\
%
\subsection{Massive Vectors in Torsionless MAG}
We proceed\footnote{Here in particular, but everywhere in general, typos are hardly missed while managing the appearance of the formulas. The raw output is available upon request.} now by profiling the vulnerabilities of the MAG program through explicit computations in ghost-free models adorned with uncontrolled interactions. Simplification in illustrating the results is the main criterion behind the choice of our Lagrangians and we hope to convince the reader, as we have convinced ourselves by working through a multitude of less presentable and more tortuous examples, over the general character of the results. 
We will consider two paradigmatic scenarios, moving from methods and results of \cite{Percacci:2020ddy,Marzo:2021esg,Barker:2024ydb}. At first, we explore the case where interactions are necessarily generated by the particular choice of the quadratic Lagrangian, the latter dictated by the absence of ghosts and tachyons. Then, we will consider the simpler scenario characterized by a particular tuning that eliminates the automatic generation of vertices. This example will possibly provide a more direct appreciation of the role of symmetries in designing consistent interactions.  \\
The building blocks of our survey are a subgroup of the invariants studied in \cite{Percacci:2020ddy}, to which we refer and direct the reader. 
In our search for simple models that are representative of the general radiative behaviour, limitation to the fewest number of couplings is favourable. We will begin with the following combination  
\begin{align} \label{ActionTLessMAG}
	& S_2[g,A]=-\frac{1}{2}\int\mathrm{d}^4x\sqrt{-g}\bigg[-a_0 F + F{^{\mu \nu \rho \sigma}} \Big(
h_1 F{_{\mu \nu \rho \sigma}} + h_2 F{_{\mu \nu \sigma \rho}} + h_3 F{_{\rho \sigma \mu \nu }} + h_4 F{_{\mu  \rho  \nu \sigma}}
 \Big) + \nn \\
 & + F^{13}{^{\mu\nu}} \Big( h_7 F^{13}{_{\mu\nu}} + h_8 F^{13}{_{\nu\mu}} \Big) +  F^{14}{^{\mu\nu}} \Big( h_9 F^{14}{_{\mu\nu}} + h_{10} F^{14}{_{\nu\mu}} \Big)  + F^{14}{^{\mu\nu}} \Big( h_{11} F^{13}{_{\mu\nu}} + h_{12} F^{13}{_{\nu\mu}} \Big)
    \bigg],
\end{align}
which source the propagation (and interactions) for the graviton as well as for a large fraction of the states in eq.~\ref{ParticleDecT}. We seek to highlight the effects solely due to the states carried by the dynamical connection $A_{\sigma \,\,\, \mu}^{\,\,\,\,\rho}$, thus decoupling the purely gravitational counterpart described by an Einstein-Hilbert term. This first step is facilitated, but not generally accomplished, by a \emph{post-Riemannian} expansion $A_{\sigma \,\,\, \mu}^{\,\,\,\,\rho}  = \Gamma^{\nu}_{\,\,\sigma \mu} + K_{\sigma \,\,\, \mu}^{\,\,\,\,\rho}$, so that the shifting symmetry $\delta A_{\mu \,\,\, \sigma}^{\,\,\,\,\rho}(x) = \partial_{\mu} \partial_{\sigma} \xi^{\rho}(x) $ is now channelled through the Levi-Civita connection $\Gamma^{\nu}_{\,\,\sigma \mu}$. Such a redefinition does not automatically prevent quadratic mixing terms between the graviton field and the newly introduced rank-3 field $K_{\sigma \,\,\, \mu}^{\,\,\,\,\rho}$ (often referred to as the \emph{distortion tensor}). Again, we include in our simplifying restrictions the absence of quadratic mixing of this type, which would only introduce an unnecessary technical nuisance. 

In a further attempt to simplify our presentation, we limited our study to only one dimensionful parameter $a_0$. Consequently, the mass $m_V$ of the vector state we intend to profile will be proportional to Planck mass $m_V^2 \sim \lambda \, a_0$. This is at odds with the EFT approach we are adopting, given that the Planck scale marks, via the dominant Einstein-Hilbert term, the boundaries of validity of any reliable, perturbative use of eq.~\ref{ActionTLessMAG}. We comment on this stressing again that the presence of a unique dimensionful term in eq.~\ref{ActionTLessMAG} is only adopted to provide a convenient showcase. A complete disentanglement of the vector masses from the Planck scale is achievable at the price of adding, for instance, quadratic combinations of the \emph{metricity} tensor $Q_{\lambda \mu \nu} \equiv - \partial_{\lambda}g_{\mu \nu} + 2 A_{\lambda\,(\mu\,\nu)}$. Moreover, even in the absence of further dimensionful parameters, the coupling responsible for the correlation, $m_V^2 \sim \lambda \, a_0$, can help to parametrically separate the two scales, a manoeuvre that is particularly harmless when such coupling only appears in the mass formula and not in the definition of the vertices. 

Once accounted for such requirements, we are in a position to only focus on the dynamic of a self-interacting high-rank tensor $K_{\sigma \,\,\, \mu}^{\,\,\,\,\rho}$, seeking the closest resemblance with the main stages of the presented treatment of interacting Proca theories. 
In particular, given our focus on the impact of radiative corrections over propagation, we need an "updated" version of eq.~\ref{RenAction} and eq.~\ref{RenActionEff}. \\
As self-evident from the post-Riemann expansion, we need to consider a subset of operators of \emph{quadratic gravity plus $K_{\sigma \,\,\, \mu}^{\,\,\,\,\rho}$ } ($G^2K$)
\begin{align} \label{LQ2K}
    \mathcal{S}_{\scriptscriptstyle {G^2K}} =   \mathcal{S}_{\scriptscriptstyle {g}} + \mathcal{S}_{\scriptscriptstyle {\nabla^2}} + \mathcal{S}_{\scriptscriptstyle {K^2}} + \mathcal{S}_{\scriptscriptstyle {R \nabla K}} + \mathcal{S}_{\scriptscriptstyle {K^{3}}} + \mathcal{S}_{\scriptscriptstyle {R K^2}}  + \mathcal{S}_{\scriptscriptstyle {K^{4}}}  + \cdots \, 
\end{align}
In eq.~\ref{LQ2K}, the terms in 
\begin{align}
  &  \mathcal{S}_{\scriptscriptstyle {g}} = \int d^4 x \sqrt{-g} \Bigg[  
\alpha_0 R + \beta_{1} R^2 + \beta_{2} R_{\mu \nu } R^{\mu \nu }
  \Bigg]\, ,
\end{align}
and
\begin{align}
  &  \mathcal{S}_{\scriptscriptstyle {R \nabla  K}} = \int d^4 x \sqrt{-g} \Bigg[  
  \eta_{1}^{} \cdot R^{\alpha \mu} \nabla_{\mu}K_{\alpha }{}^{\beta }{}_{\beta } + \eta_{3}^{} \cdot R \nabla_{\mu}K^{\alpha \mu}{}_{\alpha } + \eta_{5}^{} \cdot R^{\alpha \mu} \nabla_{\beta }K_{\alpha \mu}{}^{\beta } + \eta_{6}^{} \cdot R^{\alpha \mu} \nabla_{\beta }K_{\alpha }{}^{\beta }{}_{\mu} 
  \Bigg]\, ,
\end{align}
include now the Riemann curvature $R_{\mu \nu \,\, \sigma}^{\,\,\,\,\,\,\rho}$, attained from the Levi-Civita connection $\Gamma_{\,\,\mu \nu}^{\rho}$. These two components of $\mathcal{S}_{\scriptscriptstyle {G^2K}} $ are directly connected to the (self-)interactions of the metric fluctuation $g_{\mu \nu}(x) - \bar{g}_{\mu \nu}(x) = k_{\scriptscriptstyle G} \, h_{\mu \nu}(x) $,  $\bar{g}_{\mu \nu}(x)$ being a classical background. Moreover, they will be parametrically dependent on the corresponding expansion coupling $k_{\scriptscriptstyle G}^2 = -4/a_0$ (eq.~\ref{ActionTLessMAG}). 

The two-point Green functions for the field $K_{\sigma \,\,\, \mu}^{\,\,\,\,\rho}$, analogue of eq.~\ref{amplitude}, are instead, up to two-derivatives, encoded within the remaining factors 
\begin{align}
  &  \mathcal{S}_{\scriptscriptstyle {\nabla^2}} = \int d^4 x \sqrt{-g} \Bigg[ \zeta_{1}^{} \cdot \nabla_{\mu}K_{\beta }{}^{\nu}{}_{\nu} \nabla^{\beta }K^{\alpha }{}_{\alpha }{}^{\mu} + \zeta_{2}^{} \cdot \nabla_{\mu}K^{\nu}{}_{\beta \nu} \nabla^{\beta }K^{\alpha }{}_{\alpha }{}^{\mu} + \zeta_{3}^{} \cdot \nabla_{\beta }K_{\mu}{}^{\nu}{}_{\nu} \nabla^{\beta }K^{\alpha }{}_{\alpha }{}^{\mu} + \zeta_{4}^{} \cdot \nabla_{\beta }K^{\nu}{}_{\mu \nu} \nabla^{\beta }K^{\alpha }{}_{\alpha }{}^{\mu} + \nn \\  
    & +  \zeta_{5}^{} \cdot \nabla_{\mu}K^{\nu}{}_{\beta \nu} \nabla^{\beta }K^{\alpha \mu}{}_{\alpha } + \zeta_{6}^{} \cdot \nabla_{\beta }K^{\nu}{}_{\mu \nu} \nabla^{\beta }K^{\alpha \mu}{}_{\alpha } + \zeta_{7}^{} \cdot \nabla_{\mu}K^{\alpha \mu \beta } \nabla_{\nu}K_{\alpha }{}^{\nu}{}_{\beta } + \zeta_{8}^{} \cdot \nabla_{\alpha }K^{\alpha \mu \beta } \nabla_{\nu}K_{\mu \beta }{}^{\nu} + \nn \\ 
    & + \zeta_{9}^{} \cdot \nabla^{\beta }K^{\alpha }{}_{\alpha }{}^{\mu} \nabla_{\nu}K_{\mu \beta }{}^{\nu} + \zeta_{10}^{} \cdot \nabla^{\beta }K^{\alpha \mu}{}_{\alpha } \nabla_{\nu}K_{\mu \beta }{}^{\nu} + \zeta_{11}^{} \cdot \nabla_{\alpha }K^{\alpha \mu \beta } \nabla_{\nu}K_{\mu}{}^{\nu}{}_{\beta } + \zeta_{14}^{} \cdot \nabla_{\alpha }K^{\alpha \mu \beta } \nabla_{\nu}K_{\beta \mu}{}^{\nu} + \nn \\ 
    & + \zeta_{15}^{} \cdot \nabla^{\beta }K^{\alpha }{}_{\alpha }{}^{\mu} \nabla_{\nu}K_{\beta \mu}{}^{\nu} + \zeta_{16}^{} \cdot \nabla^{\beta }K^{\alpha \mu}{}_{\alpha } \nabla_{\nu}K_{\beta \mu}{}^{\nu} + \zeta_{24}^{} \cdot \nabla_{\nu}K_{\alpha \mu \beta } \nabla^{\nu}K^{\alpha \mu \beta } + \zeta_{25}^{} \cdot \nabla_{\nu}K_{\alpha \beta \mu} \nabla^{\nu}K^{\alpha \mu \beta }  \Bigg]\, ,
\end{align}
and
\begin{align}
    &  \mathcal{S}_{\scriptscriptstyle {K^2}} = \int d^4 x \sqrt{-g} \Bigg[
\lambda_{1}^{} \cdot K_{\alpha \mu \beta } K^{\alpha \mu \beta } + \lambda_{2}^{} \cdot K_{\alpha \beta \mu} K^{\alpha \mu \beta } + \lambda_{3}^{} \cdot K^{\alpha }{}_{\alpha }{}^{\mu} K_{\mu}{}^{\beta }{}_{\beta } + \lambda_{4}^{} \cdot K^{\alpha }{}_{\alpha }{}^{\mu} K^{\beta }{}_{\mu \beta } + \lambda_{5}^{} \cdot K^{\alpha \mu}{}_{\alpha } K^{\beta }{}_{\mu \beta }
    \Bigg]\, .
\end{align}
These last two objects form the target of our radiative assessment shedding light over the detuning of the kinetic term, the corresponding ghost resurgence and the generic obstacles in providing a predictive EFT approach.  \\
%
\subsubsection{Model 1}
We get to our first example by purposely repeating common steps adopted in order to select candidate MAG models. 
We proceed algorithmically by imposing to eq.~\ref{ActionTLessMAG} the absence, upon post-Riemann expansion, of the mixing terms $\mathcal{S}_{\scriptscriptstyle {R \nabla  K}}$ and the requirement of propagating only a healthy vector. Through established techniques involving Spin-Parity operators (SPO) and quadratic decomposition, we can easily project eq.~\ref{ActionTLessMAG} onto its $G^2 K$ representation finding 
\begin{align}
\left[
\begin{aligned}
& \zeta_1 = -\frac{h_8}{2}\, , \hspace{0.3cm} \zeta_2 = -\frac{h_{12}}{2}\, ,  \hspace{0.3cm} \zeta_3 = -\frac{h_{7}}{2}\, ,  \hspace{0.3cm} \zeta_4 = -\frac{h_{11}}{2}\, ,   \hspace{0.3cm} \zeta_5 = -\frac{h_{10}}{2}\, ,  \hspace{0.3cm} \zeta_6 = -\frac{h_{9}}{2}\,   \nn \\
& \zeta_7 = -\frac{h_{3} + h_{4} + h_{7} +h_{8}}{2}\, ,  \hspace{0.3cm} \zeta_8 = -\frac{h_{3} + h_{10}}{2} + h_{2}\, ,   \hspace{0.3cm} \zeta_9 = -\frac{h_{12}}{2} + h_{7} + h_{8}  \nn \\
& \zeta_{10} = - \frac{h_{11} + h_{12}}{2} + h_{10} \, ,  \hspace{0.3cm} \zeta_{11} = - \frac{h_{11} + h_{12}}{2} + h_{3} + h_{4} \, , \, \nn \\
& \zeta_{14} = h_1 - \frac{h_{9}}{2}  \, ,  \hspace{0.3cm}  \zeta_{15} = \frac{h_{11}}{2}  \, ,  \hspace{0.3cm} \zeta_{16} = h_{9}   \, ,  \hspace{0.3cm} \zeta_{24} = - h_{1}   \, ,  \hspace{0.3cm} \zeta_{25} = - h_{2} - \frac{h_{4}}{2}  \,  \nn \\ 
& \lambda_1 = \lambda_3 = \lambda_5 = 0\, ,  \hspace{0.3cm}  \lambda_2 = - \lambda_4 = \frac{a_0}{2} \,  \nn 
\end{aligned}
\right]
\end{align}
plus
\begin{align}
\left[
\begin{aligned}
& \eta_1 = -\frac{1}{2}\left(h_{11}  + h_{12} \right) + h_7 + h_8\, ,  \hspace{0.3cm} \eta_3 = \frac{1}{4}\left(h_{11} + h_{12} - 2 (h_9 + h_{10}) \right) \, ,  \hspace{0.3cm} \eta_5 = 2 \left(h_1 - h_2 + h_3 \right) + h_4 + h_9 -\frac{1}{2}\left( h_{11} + h_{12}\right)\,  \nn \\
& \eta_{6} = - 2 \left(h_1-h_2+h_3 \right) - h_4 - h_7 - h_8 +\frac{1}{2}\left(h_{11} + h_{12}\right)\,  \nn \\
& \alpha = \frac{a_0}{2}\, ,  \hspace{0.3cm} \beta_1 = - \frac{1}{2}\left(h_{7} + h_{8} + h_{9} + h_{10} - h_{11} - h_{12}  \right)  + 2 \left(- h_1 + h_2 - h_3\right) - h_4 \, ,  \hspace{0.3cm} \beta_2 =  \frac{1}{2}\left(h_{1} - h_{2} + h_{3} + h_{4}\right)
\end{aligned}
\right]
\end{align}
and, as we will see, further self-interactions. Before turning to those, which determine the relevant Feynman rules, we set the mixing coefficients to zero, yielding 
\begin{align}
&  h_{1} = (h_2 - h_3) - \frac{h_4}{2} \, , \,\,  h_{10} = h_7 + h_8- h_{9} \, , \,\,  h_{11} = 2 (h_7 + h_8) - h_{12} . \, \nn %
\end{align}
We then, through a familiar process \cite{Percacci:2020ddy,Marzo:2021esg, Barker:2024ydb}, detune the unwanted propagating states. For instance, to remove the single particle poles in the propagator connected to the spin-3 sector, we compute the determinant of the corresponding spin/parity matrix $a_{i j }^{\left\{3 -\right\}}$. In this particular case, such a matrix has one single element giving 
\begin{align}
 Det[a^{\left\{3 -\right\}}_{i,j}] = a^{\left\{3 -\right\}}_{1,1} \propto q^2 (h_3 - 2 h_2) - \frac{a_0}{2}\, ,     \nn
\end{align}
which will \emph{not} contribute to the propagation once $h_3 = 2 h_2$ is imposed. Notice how this does not generate a (possibly accidental) symmetry, given the presence of the non-zero $a_0$ term. 
Continuing this procedure throughout all the states in eq.~\ref{ParticleDecT}, in an attempt to select a healthy propagation for a massive vector and the graviton, we find the further constraints
\begin{align}
    h_4 = h_3 = h_2 = 0\, , \,\, h_7 = - h_8\, , \, h_{12} = - \frac{1}{3} \left( 3 + \sqrt{15} \right) h_8 \,, \, h_{12} = - \frac{1}{6} \left( 4 + \sqrt{15} \right) h_8 \, . \nn
\end{align}
After a first inessential adjustment $h_8 \rightarrow \tilde h_8 \left( 4 + \sqrt{15} \right)$ we find a propagating axial vector with mass $m_V^2=m^2_{1^+} = - 3 a_0 / \tilde h_8$. As mentioned, having the interaction coupling $\tilde h_8$ as the unique proportionality factor between the vector mass and the Planck scale is a red flag for an EFT interpretation of the complete MAG system. As also mentioned, parametrically decoupling the two scales is feasible at the expense of the introduction of further quadratic terms. By doing so we would gain in rigour what we lose in clarity. We opt therefore to keep this simple one-parameter theory that displays the generic features of one-loop MAG. Moreover, given the successful decoupling of the quadratic terms $\mathcal{S}_{\scriptscriptstyle {R \nabla  K}}$, we can isolate our study to that of an EFT for the sole self-interacting distortion $K_{\mu \nu \rho}$ hiding the large scale $a_0$ in the definition of the mass $m^2_{1^+} = - 3 a_0 / \tilde h_8$. 
Upon computation of the saturated propagator, we find that the resulting quadratic Lagrangian is free from ghosts and tachyons if $\tilde h_8 > 0$ and $a_0 < 0$, the latter required by the healthy gravitational propagation.
\\
As anticipated, this model naturally provides the dynamics of the distortion tensor with a selection of cubic and quartic self-interaction proportional to the independent coupling $\tilde h_8$. In other words, non zero $\tilde h_8$ not only triggers the propagation but also the three and four-point vertices. Again, this is not dissimilar from the appearances of Yang-Mills theory where we have  
\begin{align}
&   - \frac{1}{4 g^2} F^a_{\mu \nu} F^a{}^{\mu \nu} = - \frac{1}{4 g^2} \left( \partial_{\mu} B^a_{\nu} - \partial_{\nu} B^a_{\mu} - f^{a b c } B^b_{\mu} B^c_{\nu} \right)^2 = \nn \\
&  \left[ \text{ after rescaling } B \rightarrow g B \right] = - \frac{1}{4} \left( \partial_{\mu} B^a_{\nu} - \partial_{\nu} B^a_{\mu} - g f^{a b c } B^b_{\mu} B^c_{\nu} \right)^2 \,\, . \nn
\end{align}
Similarly, we can identify the opportune small coupling of the perturbative expansion redefining $\tilde h_8 = 1/g_{\scriptscriptstyle K}^2$ and rescaling the field accordingly as
\begin{align}
  &  \tilde h_8 \left(K^2 + K^3 + K^4\right) =  \frac{1}{g_{\scriptscriptstyle K}^2} \left(K^2 + K^3 + K^4\right) = \nn \\ 
  & \left[ \text{ after rescaling } K \rightarrow g_{\scriptscriptstyle K} K \right] = K^2 + g_{\scriptscriptstyle K} K^3 + g_{\scriptscriptstyle K}^2 K^4 \,\, .    \nn
\end{align}
We stress that none of the latter manipulations are necessary for the advancement of the computation, nor affect the final results beyond the obvious relabelling. They are only justified to shape the appearance of the diagrammatic computation in a more familiar form. 
We have now the MAG counterparts of eq.~\ref{ProcaIn} for the distortion, generated by the tuning of its quadratic sector. 
We define in eq.~\ref{LQ2K} the following cubic 
\begin{align}
&  \mathcal{S}_{\scriptscriptstyle {K^3}} =  \frac{g_{\scriptscriptstyle K}}{6}\, \int d^4 x \sqrt{-g} \Bigg[ \nn \\ 
&   (-3 + \sqrt{15}) \, K_{\mu \rho }{}^{\beta } K^{\mu \nu \rho } \nabla_{\beta }K_{\nu }{}^{\sigma }{}_{\sigma } +  (3 -  \sqrt{15})\, K^{\mu \nu }{}_{\mu } K_{\nu }{}^{\rho \beta } \nabla_{\beta }K_{\rho }{}^{\sigma }{}_{\sigma } -  K_{\mu \rho }{}^{\beta } K^{\mu \nu \rho } \nabla_{\beta }K^{\sigma }{}_{\nu \sigma } + \nn \\ 
& +  K^{\mu \nu }{}_{\mu } K_{\nu }{}^{\rho \beta } \nabla_{\beta }K^{\sigma }{}_{\rho \sigma } + (3 - \sqrt{15}) \, K_{\mu \rho }{}^{\beta } K^{\mu \nu \rho } \nabla_{\nu }K_{\beta }{}^{\sigma }{}_{\sigma } +  K_{\mu \rho }{}^{\beta } K^{\mu \nu \rho } \nabla_{\nu }K^{\sigma }{}_{\beta \sigma } +  (-3 + \sqrt{15}) \, K^{\mu \nu }{}_{\mu } K_{\nu }{}^{\rho \beta } \nabla_{\rho }K_{\beta }{}^{\sigma }{}_{\sigma } + \nn \\ 
& -   K^{\mu \nu }{}_{\mu } K_{\nu }{}^{\rho \beta } \nabla_{\rho }K^{\sigma }{}_{\beta \sigma } +  K_{\mu \rho }{}^{\beta } K^{\mu \nu \rho } \nabla_{\sigma }K_{\beta \nu }{}^{\sigma } -   K^{\mu \nu }{}_{\mu } K_{\nu }{}^{\rho \beta } \nabla_{\sigma }K_{\beta \rho }{}^{\sigma } -  K_{\mu \rho }{}^{\beta } K^{\mu \nu \rho } \nabla_{\sigma }K_{\nu \beta }{}^{\sigma } +  K^{\mu \nu }{}_{\mu } K_{\nu }{}^{\rho \beta } \nabla_{\sigma }K_{\rho \beta }{}^{\sigma } \Bigg] \, , \nn \\
\end{align}
and quartic contributions
\begin{align}
&  \mathcal{S}_{\scriptscriptstyle {K^4}} =  \frac{g_{\scriptscriptstyle K}^2}{12}\, \int d^4 x \sqrt{-g} \Bigg[ 
K^{f}{}_{\sigma f} K_{\beta \rho }{}^{\sigma } K^{\mu \nu }{}_{\mu } K_{\nu }{}^{\rho \beta } -  K^{f}{}_{\sigma f} K^{\mu \nu }{}_{\mu } K_{\nu }{}^{\rho \beta } K_{\rho \beta }{}^{\sigma } -   K_{\mu \rho }{}^{\beta } K^{\mu \nu \rho } K_{\nu }{}^{\sigma f} K_{\sigma \beta f} + \nn \\ 
& + 2\, K^{\mu \nu }{}_{\mu } K_{\nu }{}^{\rho \beta } K_{\rho }{}^{\sigma f} K_{\sigma \beta f} +  K_{\beta }{}^{\sigma f} K_{\mu \rho }{}^{\beta } K^{\mu \nu \rho } K_{\sigma \nu f} - 2\, K_{\beta }{}^{\sigma f} K^{\mu \nu }{}_{\mu } K_{\nu }{}^{\rho \beta } K_{\sigma \rho f}
\Bigg] \, , \nn \\
\end{align}
which will generate the same one-loop two-point topologies of Fig.~\ref{Fig:AA}.

\subsubsection{Model 2}
The number of independent quadratic invariants of the non-Riemann curvature eq.~\ref{curvature} allows for a further scenario. It is possible to select a particular combination that is not only free of ghosts and tachyons but is also devoid of the generated cubic and quartic self-interactions in the distortion. This is the known case (for instance, see \cite{Barker:2024ydb}) obtainable from eq.~\ref{ActionTLessMAG} by setting all parameters but $a_0$, $h_7$ and $h_8$ to zero. A healthy vector is then propagating when $h_8 = -h_7 < 0 $, with positive residue and mass $m_V^2 = m_{1^-}^2 = -3 a_0/ h_7$.
While this massive vector model is not deprived of dynamics, due to its gravitational embedding, no blatant detuning is expected from the proper inclusion of the graviton propagation. This is expected to generate (gauge-dependent) corrections to the kinetic term highly suppressed by inverse powers of the Planck mass, safely pushing the sources of non-unitary propagation outside of the validity of an EFT approach. 
We want instead to continue our comparison with the standard Proca model by considering how uncontrolled self-interactions are potentially disruptive of any EFT ambition.

We extend the quadratic Lagrangian with the following terms
\begin{align} \label{mod2}
 &  \mathcal{S}_{\scriptscriptstyle {FQ^2}} =  - \frac{1}{2} \int d^4 x \sqrt{-g} \bigg[d_1 \cdot F Q_{\mu} Q^{\mu} + d_2 \cdot F \tilde Q_{\mu} \tilde Q^{\mu} \bigg] \, , 
\end{align}
where the non-metricity tensor $Q_{\lambda \mu \nu} \equiv - \partial_{\lambda}g_{\mu \nu} + 2 A_{\lambda\,(\mu\,\nu)}$ appears through the contractions $Q_{\mu} \equiv Q_{\mu \,\,\, \lambda}^{\,\,\,\,\lambda} $ and $     \tilde Q_{\mu} \equiv Q_{\lambda \,\,\, \mu}^{\,\,\,\,\lambda}$ . After post-Riemann expansion we generate the cubic vertices
\begin{align}
&  \mathcal{S}_{\scriptscriptstyle {K^3}} =  \int d^4 x \sqrt{-g} \Bigg[ \nn \\ 
& \tfrac{1}{2} \, (d_1 + 4 d_2) \, K^{\mu }{}_{\mu }{}^{\nu } K_{\nu }{}^{\rho }{}_{\rho } \nabla_{\sigma }K^{\beta }{}_{\beta }{}^{\sigma } + d_1 \, K^{\mu }{}_{\mu }{}^{\nu } K^{\rho }{}_{\nu \rho } \nabla_{\sigma }K^{\beta }{}_{\beta }{}^{\sigma } + \tfrac{1}{2} \, d_1 \, K^{\mu \nu }{}_{\mu } K^{\rho }{}_{\nu \rho } \nabla_{\sigma }K^{\beta }{}_{\beta }{}^{\sigma } + \nn \\ 
& - (\tfrac{1}{2} d_1 + 2 d_2)\, K^{\mu }{}_{\mu }{}^{\nu } K_{\nu }{}^{\rho }{}_{\rho } \nabla_{\sigma }K^{\beta \sigma }{}_{\beta } -  d_1 \, K^{\mu }{}_{\mu }{}^{\nu } K^{\rho }{}_{\nu \rho } \nabla_{\sigma }K^{\beta \sigma }{}_{\beta } -  \tfrac{1}{2} \, d_1 \, K^{\mu \nu }{}_{\mu } K^{\rho }{}_{\nu \rho } \nabla_{\sigma }K^{\beta \sigma }{}_{\beta }
\Bigg] \, , \nn \\
\end{align}
as well as the quartic vertices
\begin{align}
&  \mathcal{S}_{\scriptscriptstyle {K^4}} =  \int d^4 x \sqrt{-g} \Bigg[ \nn \\ 
& -( \tfrac{1}{2} d_1 + 2 d_2) \, K^{\alpha}{}_{\sigma \alpha} K^{\beta }{}_{\beta }{}^{\sigma } K^{\mu }{}_{\mu }{}^{\nu } K_{\nu }{}^{\rho }{}_{\rho } + \tfrac{1}{2} \, (d_1 + 4 d_2) \, K_{\beta \alpha  \sigma } K^{\beta \sigma \alpha } K^{\mu }{}_{\mu }{}^{\nu } K_{\nu }{}^{\rho }{}_{\rho } -  d_1 \, K^{\alpha }{}_{\sigma \alpha } K^{\beta }{}_{\beta }{}^{\sigma } K^{\mu }{}_{\mu }{}^{\nu } K^{\rho }{}_{\nu \rho } + \nn \\ 
& + d_1  \, K_{\beta \alpha \sigma } K^{\beta \sigma \alpha } K^{\mu }{}_{\mu }{}^{\nu } K^{\rho }{}_{\nu \rho } -  \tfrac{1}{2} \, d_1 \, K^{\alpha }{}_{\sigma \alpha } K^{\beta \sigma }{}_{\beta } K^{\mu }{}_{\mu }{}^{\nu } K^{\rho }{}_{\nu \rho } + \tfrac{1}{2} \, d_1 \, K_{\beta \alpha \sigma } K^{\beta \sigma \alpha } K^{\mu \nu }{}_{\mu } K^{\rho }{}_{\nu \rho }
\Bigg] \, , \nn \\
\end{align}
and higher-order terms in curvature and distortion
\begin{align}
&  \mathcal{S}_{\scriptscriptstyle {R K^2}} =  \int d^4 x \sqrt{-g} \bigg[  -\tfrac{1}{2} \, (d_1 + 4 d_2) \, R \, K^{\alpha \,\, \beta}_{\,\,\, \alpha} K_{\beta \,\, \rho}^{\,\,\, \rho} - d_1 \, R \, K^{\alpha \,\, \beta}_{\,\,\, \alpha} K_{\,\, \beta  \rho}^{\rho}  - d_1 \tfrac{1}{2} \, R\, K^{\alpha \beta}_{ \,\,\,\, \alpha} K^{\rho}_{\,\,\, \beta \rho}  
\bigg] \, . \nn \\
\end{align}
Again, when concerned about the two-point one-loop corrections, we can limit our analysis to the topologies of Fig.~\ref{Fig:AA}.
%
\subsection{One-Loop results}
The investigation of the one-loop imprints of MAG models has recently witnessed a renovated interest \cite{Melichev:2023lwj,Martini:2023apm}. In these latter efforts, the non-renormalizable character of the underlying dynamics is manifested through the generation of new operators with increasing order in the derivatives. The methods and scope of our study are different. About the methods, we have already discussed at length how our concerns orbit around the \emph{modern} (as opposed to Dyson's) definition of non-renormalizability \cite{Gomis:1995jp}: new operators are generated that were set to zero, for structural reasons (unitarity, tachyon-freedom), in the starting Lagrangian. Moreover, in parallel with the presented failure of the EFT approach in more traditional Proca models Sec.~\ref{ProcaWeird}, we aim to illustrate how the newly generated (higher-order) operators scale with the mass of the propagating vector and therefore are not attenuated by inverse powers of the EFT cutoff.    

About our computational route, we have not adopted the elegant Heat-Kernel method \cite{Goroff:1985sz,Goroff:1985th,vandeVen:1991gw} opting for a diagrammatic approach and background field method (BFM). The use of the latter is of no relevance in the gauge-less examples, restricted to the distortion self-interactions, studied in this paper. Nevertheless, the BFM approach defines the standard realization of our coded routines, in particular in the process of testing them against known literature. 
We illustrate the main stages and tools in Appendix Sec.~\ref{diagrammatics} while we briefly discuss how the form of the one-loop local effective action is built from the two-point one-loop function. 
The use of the SPO, our standard tool in the decomposition of the tree-level quadratic Lagrangian, can be promptly transposed to the decomposition of the \emph{local} component of the effective action. The latter is shaped by the UV singular contribution of the defining Feynman diagrams. Therefore, the core of the matching between the singular operators generated by Fig.~\ref{Fig:AA} and the corresponding operators in eq.~\ref{LQ2K}
is obtained by comparing their spin/parity matrices, order by order in the external momenta.  
Consequently, and in parallel with our previous survey of the UV-divergent one-loop action of Sec.~\ref{ProcaWeird}, we introduce within eq.~\ref{LQ2K} the splitting 
\begin{align} \label{RenCoupling2}
& \zeta_i = \frac{1}{(4\pi)^2 \epsilon} \zeta_i^{\epsilon} + \zeta_i^{0} , \hspace{2cm} \lambda_i = \frac{1}{(4\pi)^2 \epsilon} \lambda_i^{\epsilon} + \lambda_i^{0} , 
\end{align}
where now, the 0 superscript points at the tree-level value and we let go of the (implicit, once properly accounted for) dependence over the \emph{MS-bar} renormalization scale $\mu$.
%
\subsubsection{Model 1 at NLO}
At tree-level, the quadratic components of the first model, after rescaling and redefinition of the dimensionful parameter, are matched into eq.~\ref{LQ2K} through the following choice of couplings
%
%
\begin{align}
&  {\zeta^0_1} =  \frac{1}{2} \left(\sqrt{15}-4\right), \hspace{0.3cm} {\zeta^0_2} =  \frac{1}{6} \left(\sqrt{15}-3\right),  \hspace{0.3cm}  {\zeta^0_3} = -{\zeta^0_1} ,  \hspace{0.3cm}  {\zeta^0_4} = -{\zeta^0_2},  \hspace{0.3cm} {\zeta^0_5} =  -\frac{1}{12},  \nn \\
& {\zeta^0_6} = -{\zeta^0_5} ,  \hspace{0.3cm}  {\zeta^0_7} =  0,  \hspace{0.3cm}  {\zeta^0_8} =  {\zeta^0_5},  \hspace{0.3cm}  {\zeta^0_9} = - {\zeta^0_2},  \hspace{0.3cm} {\zeta^0_{10}} =  \frac{1}{6},  \hspace{0.3cm}  {\zeta^0_{11}} =  0,  \hspace{0.3cm} {\zeta^0_{14}} =  \frac{1}{12}, \nn \\
&  {\zeta^0_{15}} =  \zeta^0_2, \hspace{0.3cm} {\zeta^0_{16}} = 
   -\frac{1}{6}, \hspace{0.3cm} {\zeta^0_{24}} =  0,  \hspace{0.3cm}  {\zeta^0_{25}} =  0, \nn \\
& {\lambda_1} =  0, \hspace{0.3cm} {\lambda_2} =  \frac{{m_V^2}}{6}, \hspace{0.3cm} {\lambda_3} =  0, \hspace{0.3cm} {\lambda_4} =  -\lambda_2, \hspace{0.3cm} {\lambda_5} = 0 \, . \nn
\end{align}
Projecting the one-loop two-point function into the singular part of eq.~\ref{LQ2K} reveals the following values (with the second power of the unique expansion parameter $g_{\scriptscriptstyle K}$ implicit, overall)
\begin{align}
& {\zeta^{\epsilon}_1} =  \frac{1}{192} \left(78793-20357 \sqrt{15}\right),  & & {\zeta^{\epsilon}_2} =  \frac{1}{192} \left(9641-2533 \sqrt{15}\right), \,\, \,\,  & {\zeta^{\epsilon}_3} =  \frac{1}{768} \left(81916 \sqrt{15}-316919\right), \nn \\ 
& {\zeta^{\epsilon}_4} =  \frac{1}{256} \left(3300 \sqrt{15}-13039\right),   & &{\zeta^{\epsilon}_5} =  \frac{3}{16} \left(1+3 \sqrt{15}\right),  \,\, \,\,  & {\zeta^{\epsilon}_6} =  \frac{19217-4408 \sqrt{15}}{3072}, \hspace{1.4cm} \nn \\
& {\zeta^{\epsilon}_7} =  \frac{1}{128} \left(40 \sqrt{15}-181\right),       & &{\zeta^{\epsilon}_8} =  \frac{1}{32} \left(386 \sqrt{15}-1521\right),\,\, \,\, & {\zeta^{\epsilon}_9} =  \frac{1}{192} \left(32315-8383 \sqrt{15}\right), \hspace{.2cm} \nn \\
& {\zeta^{\epsilon}_{10}} =  \frac{1}{384} \left(17249-4408 \sqrt{15}\right),& &{\zeta^{\epsilon}_{11}} =  \frac{1}{16} \left(2 \sqrt{15}-11\right), \,\, \,\, &  {\zeta^{\epsilon}_{14}} =  \frac{1}{16} \left(829-209 \sqrt{15}\right), \hspace{0.9cm} \nn \\
& {\zeta^{\epsilon}_{15}} =  \frac{1}{384} \left(16502 \sqrt{15}-63337\right), &  &{\zeta^{\epsilon}_{16}} =  \frac{1}{768} \left(10280 \sqrt{15}-41287\right), \,\, \,\, &  {\zeta^{\epsilon}_{24}} =  -\frac{5}{512} \left(88 \sqrt{15}-325\right), \hspace{0.7cm} \nn \\
& {\zeta^{\epsilon}_{25}} =  \frac{1}{256} \left(136  \sqrt{15}-487\right),   & & & \nn
\end{align}

\begin{align}
& {\lambda^{\epsilon}_1} =  \frac{1}{64} \left(180 \sqrt{15}-697\right) {m^2_V}, \,\, &  {\lambda^{\epsilon}_2} =  \frac{1}{64} \left(649-180 \sqrt{15}\right) {m^2_V}, \,\, & \hspace{1cm} {\lambda^{\epsilon}_3} =  \frac{5}{192} \left(133-36 \sqrt{15}\right) {m^2_V}, \nn \\ 
& {\lambda^{\epsilon}_4} =  \frac{5}{96} \left(36 \sqrt{15}-109\right) {m^2_V},  \,\, &  {\lambda^{\epsilon}_5} =  \frac{1}{192} \left(389-180 \sqrt{15}\right) {m^2_V} \, .& \nn
\end{align}

It is clear, as expected by the random nature of the interactions accounted, that such correction is not a renormalization of the starting tree-level action. The radiative structure highlights the presence of the missing couplings defining the high-energy quantum behavior of the model. Notice that no large mass (as expected by the dimensionality of terms investigated) comes to the rescue. As a final remark, we also point out that, again similarly to the results of Sec.~\ref{ProcaWeird}, higher orders in the momentum are also generated, that need to be merged into additional operators with higher powers in the covariant derivatives. Their particular realization lies outside of the scope of the present work. Nevertheless, for the sake of illustration, we show, arbitrarily picking the simple spin-3 sector, the form of the corresponding singular component 
\begin{align}
    a_{1,1}^{\left\{3,-\right\}} = - \frac{g_{\scriptscriptstyle K}^2}{(4 \pi)^2 \, \epsilon} \left[ \frac{3}{2} m_V^2 + \frac{9}{128}\left(8 \sqrt{15} - 31 \right) \left(\frac{7}{6}  q^2 +  \frac{q^4}{m_V^2} \right)\right] \, , 
\end{align}
highlighting once again how also higher-order poles scale with the \emph{light} mass of the theory. 
%
%
\subsubsection{Model 2 at NLO}
Finally, we take on the second scenario of added interactions. Following the same steps of the previous section we find that the simpler tree-level action is accounted for by  
\begin{align}
&  {\zeta^0_1} =  \frac{1}{2}, \hspace{0.3cm} {\zeta^0_3} = - {\zeta^0_1},  \hspace{0.3cm} {\lambda_2^0} = \frac{m_V^2}{3} , \hspace{0.3cm} \lambda^0_4 = - \lambda^0_2 , \hspace{0.3cm}  {\zeta^0_i} = 0 ,  \hspace{0.3cm}  {\lambda^0_i} = 0 \, ,\nn
\end{align}
and it is completed at one-loop by the less minimal selection
\begin{align}
    & {\zeta^{\epsilon}_7} =  {\zeta^{\epsilon}_8} =  {\zeta^{\epsilon}_9} =  {\zeta^{\epsilon}_{10}} =  {\zeta^{\epsilon}_{11}} =  {\zeta^{\epsilon}_{14}} = {\zeta^{\epsilon}_{15}} =  {\zeta^{\epsilon}_{16}} =  {\zeta^{\epsilon}_{24}} = {\zeta^{\epsilon}_{25}} = {\lambda^{\epsilon}_1} =  0 \nn
\end{align}
\begin{align}
& {\zeta^{\epsilon}_1} =  12 \, \left(73 {d_1}^2-{d_1} {d_2}+7 {d_2}^2\right), \hspace{1cm}  {\zeta^{\epsilon}_2} = - 12 \left(140 {d_1}^2+13 {d_1} {d_2}+8 {d_2}^2\right), \hspace{1cm} {\zeta^{\epsilon}_3} = -54 ({d_1}-{d_2})^2,  \nn \\
& {\zeta^{\epsilon}_4} =  108 {d_1} ({d_2}-{d_1}),  \hspace{2.6cm} {\zeta^{\epsilon}_5} =  12 \left(67 {d_1}^2+14 {d_1}{d_2} + {d_2}^2 \right), \hspace{1.6cm} {\zeta^{\epsilon}_6} = - 54 {d_1}^2, \nn %
\end{align}
\begin{align}
 & {\lambda^{\epsilon}_2} =  6 m_V^2 (4 {d_1}+{d_2}), \hspace{4cm}  {\lambda^{\epsilon}_3} =  -\frac{9}{4} m_V^2 \, \left(-8 (6 {d_1}+1) {d_2}+3 {d_1} (8 {d_1}+1)+24 {d_2}^2\right), \nn \\
 & {\lambda^{\epsilon}_4} =  -\frac{3}{2} m_V^2 \, ({d_1} (72 {d_1}-72 {d_2}+19)+4 {d_2}), \hspace{0.95cm} {\lambda^{\epsilon}_5} =  \frac{9}{4} (1-24 {d_1}) {d_1} m_V^2 \, .\nn
\end{align}
Again, the inclusion of the simple, yet unconstrained interactions of eq.~\ref{mod2} reveals that the candidate quantum theory needs a larger selection of independent couplings than those initially considered \emph{already at the two-derivative order}. Moreover, similarly to the previous example, higher-order terms carrier of higher-order poles also show up, for instance in the sector
\begin{align}
    a_{2,2}^{\left\{0,+\right\}} =  - \frac{1}{(4 \pi)^2 \, \epsilon} \left[ \cdots  + {54} \left(4 d_1 + d_2 \right)^2 \, \left( \frac{q^4}{m_V^2} - \frac{q^6}{6\, m_V^4} \right)\right] \, , 
\end{align}
which are not weakened by any inverse power of a large scale.
\section{Conclusions}
Selecting a healthy particle content in MAG is a fundamental prerequisite in revealing its phenomenological signatures. By definition, the scales so defined live beneath the cut-off scale and represent the relevant, infrared part of the spectrum. Contrarily, above the cut-off scale $m_{\Lambda}$, heavier ghostly states are allowed to appear without presenting any threat to the EFT's ability to produce coherent predictions with full control over the neglected higher-orders terms of the perturbative expansion.

This is the case, for instance, of the multi-derivative effective terms radiatively generated in the EFT treatment of Quantum Gravity ($\sim R^2$) or, as previously discussed, massive Abelian vectors (as well as of any similar attempt to build high-rank EFTs). The caveat is that the poles introduced by these new operators only signal the breakdown of the $E/m_{\Lambda}$ expansion and a proper resummation (or an altogether new UV complete theory) takes place.

As explicitly seen, and already anticipated in \cite{Gomis:1995jp}, this expectation is rarely satisfied in the presence of fields of rank $\geq 1$ unless a consistent use of symmetries is adopted, supporting the correct scaling behaviour of the pathological, radiatively generated, operators.  

We leave the present investigation, which forms the first of two parts concerning the subject of Phenomenological Lagrangians in MAG, with the negative results of the previous section and with precise profiling of their causes. We have shown how the lack of a homogeneous relation between the structure of the propagator and the added interactions, punishes our attempts to build a predictive EFT. Nevertheless, in parallel with our analysis of EFT for the traditional case of interacting rank-1 Proca theories, possible predictive MAG models are at reach, once structural symmetries are properly included. This is the subject of the forthcoming second part of our investigation.\\

\noindent \textbf{Acknowledgement.} 
We express our gratitude to William Barker, Marco Piva and Sebastian Zell for their precious comments and suggestions. 
Feynman diagrams are drawn using \texttt{TikZ} \cite{Ellis:2016jkw}. Computations in vector rank-1 models have also been checked with the help of \texttt{FeynRules} \cite{Christensen:2008py,Alloul:2013bka}. Credit for the ghost's silhouette of figure \ref{fig:ghosts} goes to utent Schmidsi of the online image stock \texttt{Pixabay}.  \\
This work was supported by the Estonian Research Council grants PRG1677, RVTT3, RVTT7, and the CoE program TK202 ``Fundamental Universe''.

\bibliography{main}

\newpage

\appendix

\section{Diagrammatics for high-rank fields} \label{diagrammatics}
The challenging task of profiling, at one-loop, the dynamic of gravity plus a rank-3 tensor (early studies can be found in \cite{YuBaurov:2018pyj,Buchbinder:1985jc,Kalmykov:1994fm}, with the recent \cite{Melichev:2023lwj,Baldazzi:2021kaf}) demands the development of appropriate tools to tame the intricate index structure. Different methods can be adopted for this task. The study of the gravitational system alone has triggered the development, in the past years, of techniques circumventing the direct computation of Feynman diagrams. Among these, we record the early attempts to profile the gravity one and two-loop UV divergences via the Heat-Kernel method \cite{Goroff:1985sz,Goroff:1985th,vandeVen:1991gw}, or through pre-compiled counterterm formulas \cite{tHooft:1974toh,Pronin:1996rv}.
In this paper, we rely on the direct computation of one-loop diagrams in BFM. No mainstream tools are available for any of the steps required and, consequently, we have developed our own computational chain. The central problem of manipulating the large expressions is managed by the use of the symbolic language \texttt{FORM} \cite{Vermaseren:2000nd,Ruijl:2017dtg,Kuipers:2012rf}. This established the unmovable core of our computation, while the less demanding side tasks could easily be tackled with a multitude of alternative tools. 
\begin{figure}[H] \label{SchemeComp}
  \centering
  \includegraphics[width=1\textwidth]{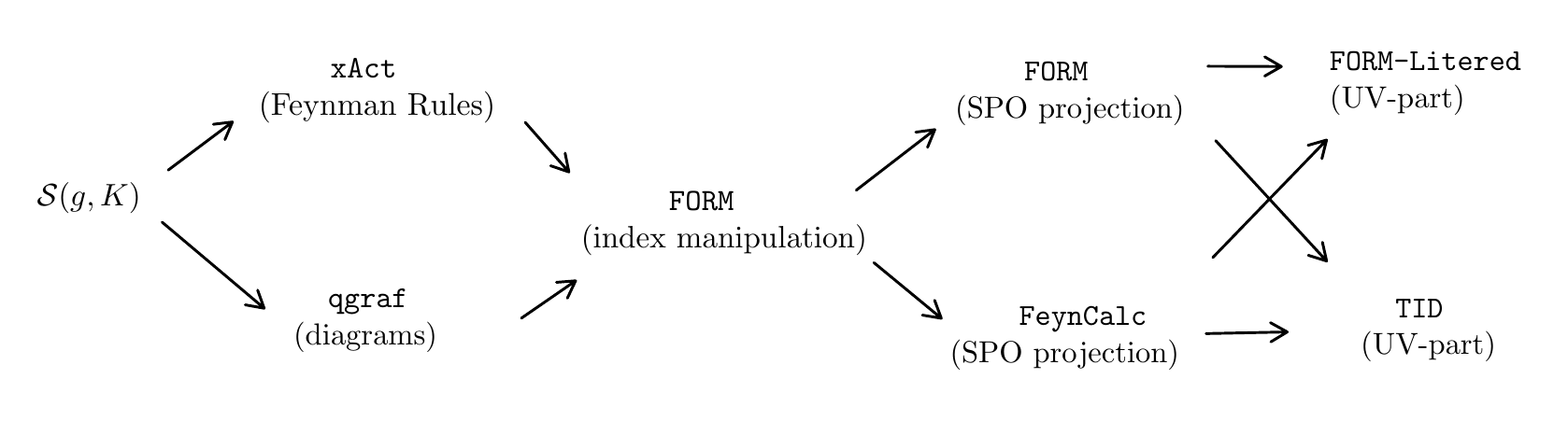} 
  \caption{Work chain of the loop computation}
  \end{figure}
The resulting work chain is schematized in \ref{SchemeComp}. We used \texttt{qgraf} \cite{Nogueira:1991ex} to encode the needed one-loop topologies. A \texttt{FORM} routine dresses such topologies with the Feynman rules obtained with the help of tensor manipulations in \texttt{xAct} \cite{Brizuela:2008ra}. Alternative ways are then explored to consistently control the final result. Projections to scalar form factors are handled by appropriate operators both within \texttt{FORM} as well as via \texttt{FeynCalc} \cite{Mertig:1990an,Shtabovenko:2016sxi,Shtabovenko:2020gxv}. Similarly, the integral scalar reduction is built within our \texttt{FORM} code, using simplifying reduction formulas found with the help of \texttt{Litered}\footnote{Admittedly, the use of such a powerful system is overkill for the simple denominator structure of a one-loop computation. Nevertheless, on top of providing alternative ways to check consistency, it prepares our routines for future, more demanding applications.} \cite{Lee:2012cn,Lee:2013mka}. Alternatively, such last steps can be handled by making the \texttt{FORM} output readable in \texttt{FeynCalc}, and using one-loop routines therein. 
Besides internal self-consistency, the strongest dictated by diffeomorphism invariance of the final result, we tested our chain against the known cases of matter-gravity systems \cite{tHooft:1974toh,Barvinsky:1983vpp} as well as the two-parameter gauge dependence of one-loop gravity \cite{Capper:1983fu}.
As mentioned, to reconstruct the form of the one-loop quadratic action we take advantage of the same projector machinery developed to tackle the spectral problem \cite{Percacci:2020ddy,Marzo:2021esg,Marzo:2021iok,Barker:2024ydb}.

\section{The propagators in {\bf \texttt{FORM}}}
To highlight the intricate structure of the main building blocks for typical high-rank models, we show the coded  \texttt{FORM} input of the propagators. Vertices are typically oversized for page display, necessitating storage in the range of kilobytes.    

\subsection{Model 1}

\adjustbox{max width=0.6\textwidth}{
\begin{minipage}{\textwidth}
\hspace*{-1cm} %
\begin{verbatim}

    id once prop([KK],mu1?,mu2?,mu3?,mu4?,mu5?,mu6?,q1?) =
    (-i_/2)*((3*d_(mu2, mu5)*d_(mu4, mu3)*d_(mu6, mu1))/MK2 - (3*d_(mu2, mu4)*d_(mu5, mu3)*d_(mu6, mu1))/MK2 
    - ((-7 + 2*sqrt_(15))*d_(mu2, mu3)*d_(mu5, mu4)*d_(mu6, mu1))/MK2 - (3*d_(mu1, mu5)*d_(mu4, mu3)*d_(mu6, mu2))/MK2 
    - (3*d_(mu1, mu4)*d_(mu5, mu3)*d_(mu6, mu2))/MK2 + ((-24 + 7*sqrt_(15))*d_(mu1, mu3)*d_(mu5, mu4)*d_(mu6, mu2))/MK2 
    - (3*d_(mu1, mu5)*d_(mu2, mu4)*d_(mu6, mu3))/MK2 + (3*d_(mu1, mu4)*d_(mu2, mu5)*d_(mu6, mu3))/MK2 
    - ((-7 + 2*sqrt_(15))*d_(mu1, mu2)*d_(mu5, mu4)*d_(mu6, mu3))/MK2 + ((-24 + 7*sqrt_(15))*d_(mu1, mu5)*d_(mu2, mu3)*d_(mu6, mu4))/MK2
    - (6*(-15 + 4*sqrt_(15))*d_(mu1, mu3)*d_(mu2, mu5)*d_(mu6, mu4))/MK2 + ((-24 + 7*sqrt_(15))*d_(mu1, mu2)*d_(mu5, mu3)*d_(mu6, mu4))/MK2 
    - ((-7 + 2*sqrt_(15))*d_(mu1, mu4)*d_(mu2, mu3)*d_(mu6, mu5))/MK2 + ((-24 + 7*sqrt_(15))*d_(mu1, mu3)*d_(mu2, mu4)*d_(mu6, mu5))/MK2
    - ((-7 + 2*sqrt_(15))*d_(mu1, mu2)*d_(mu4, mu3)*d_(mu6, mu5))/MK2 + ((-3 + sqrt_(15))*d_(mu5, mu4)*d_(mu6, mu3)*q1(mu1)*q1(mu2))/MK2^2 
    - (3*(-4 + sqrt_(15))*d_(mu5, mu3)*d_(mu6, mu4)*q1(mu1)*q1(mu2))/MK2^2 + ((-3 + sqrt_(15))*d_(mu4, mu3)*d_(mu6, mu5)*q1(mu1)*q1(mu2))/MK2^2
    - (2*(-3 + sqrt_(15))*d_(mu5, mu4)*d_(mu6, mu2)*q1(mu1)*q1(mu3))/MK2^2 + (6*(-4 + sqrt_(15))*d_(mu2, mu5)*d_(mu6, mu4)*q1(mu1)*q1(mu3))/MK2^2
    - (2*(-3 + sqrt_(15))*d_(mu2, mu4)*d_(mu6, mu5)*q1(mu1)*q1(mu3))/MK2^2 + ((-3 + sqrt_(15))*d_(mu5, mu4)*d_(mu6, mu1)*q1(mu2)*q1(mu3))/MK2^2
    - (3*(-4 + sqrt_(15))*d_(mu1, mu5)*d_(mu6, mu4)*q1(mu2)*q1(mu3))/MK2^2 + ((-3 + sqrt_(15))*d_(mu1, mu4)*d_(mu6, mu5)*q1(mu2)*q1(mu3))/MK2^2
    + (2*(-4 + sqrt_(15))*d_(mu2, mu3)*d_(mu6, mu5)*q1(mu1)*q1(mu4))/MK2^2 + ((27 - 7*sqrt_(15))*d_(mu1, mu3)*d_(mu6, mu5)*q1(mu2)*q1(mu4))/MK2^2
    + (2*(-4 + sqrt_(15))*d_(mu1, mu2)*d_(mu6, mu5)*q1(mu3)*q1(mu4))/MK2^2 + ((27 - 7*sqrt_(15))*d_(mu2, mu3)*d_(mu6, mu4)*q1(mu1)*q1(mu5))/MK2^2
    + (3*(-31 + 8*sqrt_(15))*d_(mu1, mu3)*d_(mu6, mu4)*q1(mu2)*q1(mu5))/MK2^2 + ((27 - 7*sqrt_(15))*d_(mu1, mu2)*d_(mu6, mu4)*q1(mu3)*q1(mu5))/MK2^2
    + ((-3 + sqrt_(15))*d_(mu2, mu3)*d_(mu6, mu1)*q1(mu4)*q1(mu5))/MK2^2 - (3*(-4 + sqrt_(15))*d_(mu1, mu3)*d_(mu6, mu2)*q1(mu4)*q1(mu5))/MK2^2
    + ((-3 + sqrt_(15))*d_(mu1, mu2)*d_(mu6, mu3)*q1(mu4)*q1(mu5))/MK2^2 + (2*(-4 + sqrt_(15))*d_(mu2, mu3)*d_(mu5, mu4)*q1(mu1)*q1(mu6))/MK2^2
    + ((27 - 7*sqrt_(15))*d_(mu1, mu3)*d_(mu5, mu4)*q1(mu2)*q1(mu6))/MK2^2 + (2*(-4 + sqrt_(15))*d_(mu1, mu2)*d_(mu5, mu4)*q1(mu3)*q1(mu6))/MK2^2
    - (2*(-3 + sqrt_(15))*d_(mu1, mu5)*d_(mu2, mu3)*q1(mu4)*q1(mu6))/MK2^2 + (6*(-4 + sqrt_(15))*d_(mu1, mu3)*d_(mu2, mu5)*q1(mu4)*q1(mu6))/MK2^2
    - (2*(-3 + sqrt_(15))*d_(mu1, mu2)*d_(mu5, mu3)*q1(mu4)*q1(mu6))/MK2^2 + ((-3 + sqrt_(15))*d_(mu1, mu4)*d_(mu2, mu3)*q1(mu5)*q1(mu6))/MK2^2
    - (3*(-4 + sqrt_(15))*d_(mu1, mu3)*d_(mu2, mu4)*q1(mu5)*q1(mu6))/MK2^2 + ((-3 + sqrt_(15))*d_(mu1, mu2)*d_(mu4, mu3)*q1(mu5)*q1(mu6))/MK2^2)
    - (((3*i_)/2)*den(q1, MK2)*(MK2*d_(mu5, mu3)*d_(mu6, mu2)*q1(mu1)*q1(mu4) - MK2*d_(mu2, mu5)*d_(mu6, mu3)*q1(mu1)*q1(mu4)
    - MK2*d_(mu2, mu5)*d_(mu6, mu1)*q1(mu3)*q1(mu4) + MK2*d_(mu1, mu5)*d_(mu6, mu2)*q1(mu3)*q1(mu4) + d_(mu6, mu3)*q1(mu1)*q1(mu2)*q1(mu4)*q1(mu5)
    - 2*d_(mu6, mu2)*q1(mu1)*q1(mu3)*q1(mu4)*q1(mu5) + d_(mu6, mu1)*q1(mu2)*q1(mu3)*q1(mu4)*q1(mu5) - MK2*d_(mu2, mu5)*d_(mu4, mu3)*q1(mu1)*q1(mu6)
    + MK2*d_(mu2, mu4)*d_(mu5, mu3)*q1(mu1)*q1(mu6) + MK2*d_(mu1, mu5)*d_(mu2, mu4)*q1(mu3)*q1(mu6) - MK2*d_(mu1, mu4)*d_(mu2, mu5)*q1(mu3)*q1(mu6)
    - 2*d_(mu5, mu3)*q1(mu1)*q1(mu2)*q1(mu4)*q1(mu6) + 4*d_(mu2, mu5)*q1(mu1)*q1(mu3)*q1(mu4)*q1(mu6) - 2*d_(mu1, mu5)*q1(mu2)*q1(mu3)*q1(mu4)*q1(mu6)
    + d_(mu4, mu3)*q1(mu1)*q1(mu2)*q1(mu5)*q1(mu6) - 2*d_(mu2, mu4)*q1(mu1)*q1(mu3)*q1(mu5)*q1(mu6) 
    + d_(mu1, mu4)*q1(mu2)*q1(mu3)*q1(mu5)*q1(mu6)))/MK2^2;


\end{verbatim}
\end{minipage}
}
\\
where MK2 stands for the square of the unique propagator mass and the function den$(q,m^2)$ is $1/(q^2 - m^2)$. The label $[KK]$ helps \texttt{FORM} to discriminate between propagators and other six-index vertices like, for instance, the three graviton case. 

\subsection{Model 2}
\adjustbox{max width=0.6\textwidth}{
\begin{minipage}{\textwidth}
\hspace*{-1cm} %
\begin{verbatim}
id once prop([KK],mu1?,mu2?,mu3?,mu4?,mu5?,mu6?,q1?) = (((-3*i_)/4)*(d_(mu2, mu5)*d_(mu4, mu3)*d_(mu6, mu1) - d_(mu2, mu4)*d_(mu5, mu3)*d_(mu6, mu1) 
- d_(mu1, mu5)*d_(mu4, mu3)*d_(mu6, mu2) - d_(mu1, mu4)*d_(mu5, mu3)*d_(mu6, mu2) - d_(mu1, mu5)*d_(mu2, mu4)*d_(mu6, mu3) 
+ d_(mu1, mu4)*d_(mu2, mu5)*d_(mu6, mu3) + 2*d_(mu1, mu3)*d_(mu2, mu5)*d_(mu6, mu4)))/MK2 - ((i_/4)*den(q1, MK2)*(MK2*d_(mu2, mu3)*d_(mu5, mu4)*d_(mu6, mu1)
- 3*MK2*d_(mu1, mu3)*d_(mu5, mu4)*d_(mu6, mu2) + MK2*d_(mu1, mu2)*d_(mu5, mu4)*d_(mu6, mu3) - 3*MK2*d_(mu1, mu5)*d_(mu2, mu3)*d_(mu6, mu4) 
+ 9*MK2*d_(mu1, mu3)*d_(mu2, mu5)*d_(mu6, mu4) - 3*MK2*d_(mu1, mu2)*d_(mu5, mu3)*d_(mu6, mu4) + MK2*d_(mu1, mu4)*d_(mu2, mu3)*d_(mu6, mu5) 
- 3*MK2*d_(mu1, mu3)*d_(mu2, mu4)*d_(mu6, mu5) + MK2*d_(mu1, mu2)*d_(mu4, mu3)*d_(mu6, mu5) - d_(mu2, mu3)*d_(mu6, mu5)*q1(mu1)*q1(mu4) 
+ 3*d_(mu1, mu3)*d_(mu6, mu5)*q1(mu2)*q1(mu4) - d_(mu1, mu2)*d_(mu6, mu5)*q1(mu3)*q1(mu4) + 3*d_(mu2, mu3)*d_(mu6, mu4)*q1(mu1)*q1(mu5) 
- 9*d_(mu1, mu3)*d_(mu6, mu4)*q1(mu2)*q1(mu5) + 3*d_(mu1, mu2)*d_(mu6, mu4)*q1(mu3)*q1(mu5) - d_(mu2, mu3)*d_(mu5, mu4)*q1(mu1)*q1(mu6) 
+ 3*d_(mu1, mu3)*d_(mu5, mu4)*q1(mu2)*q1(mu6) - d_(mu1, mu2)*d_(mu5, mu4)*q1(mu3)*q1(mu6)))/MK2;


\end{verbatim}
\end{minipage}
}

\end{document}